\documentclass[12pt,eqsecnum]{article}
\usepackage[dvips]{graphicx}
\usepackage{amssymb}
\usepackage{amsmath}
\usepackage{ulem}
\usepackage[dvipsnames]{xcolor}
\usepackage{soul}
\usepackage{cite}

\makeatletter

\@addtoreset{equation}{section}
\makeatletter

\newcommand{\qed}{\hbox{\rule[-2pt]{6pt}{6pt}}}
\newcommand{\D}{{\rm d}}

\newtheorem{Prop}{Proposition}

\newcommand{\dalm}{\kern1pt\vbox{\hrule height 0.9pt\hbox{\vrule width
0.9pt\hskip 2.5pt\vbox{\vskip 5.5pt}\hskip 3pt\vrule width 0.3pt}\hrule height
0.3pt}\kern1pt}

\begin{document}

\begin{titlepage}
\vfill
\begin{flushright}
\today
\end{flushright}

\vfill
\begin{center}
\baselineskip=16pt
{\Large\bf 
Vacuum-dual static perfect fluid obeying $p=-(n-3)\rho/(n+1)$ in $n(\ge 4)$ dimensions
}
\vskip 0.5cm
{\large {\sl }}
\vskip 10.mm
{\bf Hideki Maeda} \\

\vskip 1cm
{
Department of Electronics and Information Engineering, Hokkai-Gakuen University, Sapporo 062-8605, Japan.\\
\texttt{h-maeda@hgu.jp}

}
\vspace{6pt}
\end{center}
\vskip 0.2in
\par
\begin{center}
{\bf Abstract}
\end{center}
\begin{quote}
We obtain the general $n(\ge 4)$-dimensional static solution with an $(n-2)$-dimensional Einstein base manifold for a perfect fluid obeying a linear equation of state $p=-(n-3)\rho/(n+1)$.
It is a generalization of Semiz's four-dimensional general solution with spherical symmetry and consists of two different classes.
Through the Buchdahl transformation, the class-I and class-II solutions are dual to the topological Schwarzschild-Tangherlini-(A)dS solution and one of the $\Lambda$-vacuum direct-product solutions, respectively.
While the metric of the spherically symmetric class-I solution is $C^\infty$ at the Killing horizon for $n=4$ and $5$, it is $C^1$ for $n\ge 6$ and then the Killing horizon turns to be a parallelly propagated curvature singularity.
For $n=4$ and $5$, the spherically symmetric class-I solution can be attached to the Schwarzschild-Tangherlini vacuum black hole with the same value of the mass parameter at the Killing horizon in a regular manner, namely without a lightlike massive thin-shell.
This construction allows new configurations of an asymptotically (locally) flat black hole to emerge.
If a static perfect fluid hovers outside a vacuum black hole, its energy density is negative. 
In contrast, if the dynamical region inside the event horizon of a vacuum black hole is replaced by the class-I solution, the corresponding matter field is an anisotropic fluid and may satisfy the null and strong energy conditions.
While the latter configuration always involves a spacelike singularity inside the horizon for $n=4$, it becomes a non-singular black hole of the big-bounce type for $n=5$ if the ADM mass is larger than a critical value.
\vfill
\vskip 2.mm
\end{quote}
\end{titlepage}




\tableofcontents

\newpage

\section{Introduction}

The study of static and spherically symmetric solutions with a perfect fluid in general relativity aims to find nice models representing compact objects in gravitational equilibrium such as white dwarfs or neutron stars.
Based on the Tolman-Oppenheimer-Volkoff (TOV) equation~\cite{Oppenheimer:1939ne,Tolman:1939jz}, a huge effort has been made in the history of general relativity to obtain physically reasonable solutions with a regular center that are asymptotically flat or can be attached in a regular manner at some radius to an exterior Schwarzschild vacuum region.

In the last century, exact solutions had been obtained mainly in a heuristic way by introducing nice coordinate systems or variables. (See~\cite{Delgaty:1998uy} for the results until 1998 and also Sec.~16.1 in the textbook~\cite{Stephani:2003}.)
In the 21st century, based on the earlier observations~\cite{Wyman:1949zz,bhs1987,Fodor:2000gu}, an algorithm to construct all regular static spherically symmetric perfect-fluid solutions based on the generating function has been developed without specifying an equation of state~\cite{Rahman:2001hp,Lake:2002bq,Martin:2003jc}.
This development lead to derive an infinite number of previously unknown physically interesting exact solutions as well as to establish solution-generating methods~\cite{Martin:2003jc,Boonserm:2005ni,Boonserm:2006up,Boonserm:2007zm}.
The solution space has also been studied in the dynamical systems approach with linear and polytropic equations of state~\cite{Nilsson:2000zf,Nilsson:2000zg}.

Nevertheless, a complete classification of static spherically symmetric solutions obeying a physically important linear equation of state $p=\chi\rho$ has not been achieved yet except for several particular values of $\chi$. 
(The dominant energy condition is equivalent to $\rho\ge 0$ with $-1\le \chi\le 1$ in arbitrary dimensions~\cite{Maeda:2018hqu}.)
For example, such a static solution is absent for $\chi=0$ and the general solution consists of the Schwarzschild-(A)dS and Nariai solutions for $\chi=-1$ by Birkhoff's theorem.
Ivanov studied the integrability of the field equations in detail for general $\chi$ in four dimensions and obtained a particular solution for $\chi=-1/5$ by the Buchdahl transformation from the (anti-)de~Sitter solution as a seed solution~\cite{Ivanov:2001sg}.
In this situation, Semiz classified solutions with a mass function given as a polynomial of the areal radius~\cite{Semiz:2008ny}.
Based on this result, he has recently derived the general static and spherically symmetric perfect-fluid solution for $\chi=-1/5$~\cite{Semiz:2020lxj}.
Semiz's general solution consists of two different classes and they are related to the Schwarzschild-(A)dS and Nariai $\Lambda$-vacuum solutions through the Buchdahl transformation~\cite{Semiz:2020lxj}.
Some properties of this general solution have been studied in~\cite{Fazlpour:2022rsz}.
However, there still remains room for investigation to understand these solutions and provide their physical interpretations.

In this paper, we will therefore fully investigate Semiz's general solution in a broader framework.
To be more precise, we will derive and study the general $n(\ge 4)$-dimensional static solution with an $(n-2)$-dimensional Einstein base manifold for $\chi=-(n-3)/(n+1)$.
The motivation for studying the case with negative pressure in arbitrary dimensions is not to find a model of a star-like static equilibrium configuration, but to gain insight into the nature of gravity through the analysis of exact solutions.
In fact, since a static equilibrium is realized by balancing the pressure with self-gravity, a static solution with negative $\chi$ should have a negative energy density violating the weak energy condition and/or admit a (naked) singularity.
Furthermore, a perfect fluid with negative pressure generally suffers from the hydrodynamical instability in the flat spacetime because the speed of sound $c_{\rm s}:=\sqrt{\D p/\D\rho}$ becomes pure imaginary.
Note, however, that $\chi=-1$ is an exception because such a perfect fluid is equivalent to a cosmological constant and its energy density and pressure are constant.

The present paper is organized as follows.
First, in Sec.~\ref{sec:Semiz-derive}, we will generalize Semiz's four-dimensional spherically symmetric perfect-fluid solutions to $n(\ge 4)$ dimensions with a more general base manifold of the Einstein space characterized by a curvature scalar $k(=1,0,-1)$.
Furthermore, we will show that the generalized Semiz solutions are dual to the topological Schwarzschild-Tangherlini-(A)dS solution or one of the $\Lambda$-vacuum direct-product solutions through the $n$-dimensional Buchdahl transformation.
In Sec.~\ref{sec:Semiz-property}, we will fully investigate physical and geometric properties of the $n$-dimensional Semiz class-I solution with spherical symmetry.
We will summarize our results and present concluding remarks in the final section.
Our conventions for curvature tensors are $[\nabla _\rho ,\nabla_\sigma]V^\mu ={{R}^\mu }_{\nu\rho\sigma}V^\nu$ and ${R}_{\mu \nu }={{R}^\rho }_{\mu \rho \nu }$, where Greek indices run over all spacetime indices.
The signature of the Minkowski spacetime is $(-,+,+,\cdots,+)$ and other types of indices will be specified in the main text.
We adopt the units such that $c=1$ and $\kappa_n:=8\pi G_n$, where $G_n$ is the $n$-dimensional gravitational constant.
Throughout the paper, a prime over a function denotes differentiation with respect to its argument.

\section{Semiz-class perfect-fluid solutions in $n(\ge 4)$ dimensions}
\label{sec:Semiz-derive}

In this section, we drive exact solutions to the following Einstein equations with a perfect fluid in $n(\ge 4)$ dimensions:
\begin{align}
&{\cal E}_{\mu\nu}:=G_{\mu\nu}-\kappa_nT_{\mu\nu}=0,\label{EFE-0}\\
&{T}_{\mu\nu}=(\rho+p) u_\mu u_\nu+pg_{\mu\nu}.\label{Tab-0}
\end{align}
Here $\rho$ and $p$ are the energy density and pressure of a perfect fluid, respectively, and $u^\mu$ is the normalized $n$-velocity of the fluid element satisfying $u_\mu u^\mu=-1$.

To write down the Einstein equations (\ref{EFE-0}), consider an $n(\ge 4)$-dimensional spacetime $({\cal M}^n, g_{\mu \nu })$ as a warped product of a two-dimensional Lorentzian spacetime $(M^2, g_{AB})$ and an $(n-2)$-dimensional Einstein space $(K_{(k)}^{n-2}, \gamma _{ij})$ with the following metric:
\begin{align}
\label{eq:ansatz}
\begin{aligned}
\D s^2=&g_{\mu \nu }\D x^\mu \D x^\nu \\
=&g_{AB}(y)\D y^A\D y^B +\sigma(y)^2 \gamma _{ij}(z)\D z^i\D z^j ,
\end{aligned} 
\end{align} 
where $A,B = 0, 1$ and $i,j = 2, 3,\cdots, n-1$. 
The Ricci tensor on $K_{(k)}^{n-2}$ is given by ${}^{(n-2)}{R}_{ij}=k(n-3)\gamma_{ij}$ with $k =1,0,-1$\footnote{The definition of the Einstein space is ${}^{(n-2)}R_{ij}=\lambda \gamma_{ij}$ with a constant $\lambda$, which can be set to $\lambda=k(n-3)$ without loss of generality by redefining the areal radius $\sigma(y)$.}.
For the spacetime (\ref{eq:ansatz}), the Riemann tensor is decomposed as
\begin{align}
{R}_{ABCD}&={}^{(2)}{R}_{ABCD},\\
{R}_{AiBj}&=-\sigma(D_A D_B \sigma)\gamma _{ij}, \label{eq:Riemann}\\
{R}_{ijkl}&=\sigma^2[k-(D\sigma)^2](\gamma_{ik}\gamma _{jl}-\gamma_{il}\gamma _{jk}),
\end{align}
where ${}^{(2)}{R}_{ABCD}$ and $D_A $ are the Riemann tensor and the covariant derivative on $M^2$, respectively, and $(D\sigma)^2:=g^{AB}(D_A\sigma)(D_B\sigma)$.
Also, the Einstein tensor is decomposed as
\begin{align}
G_{AB}=&-(n-2)\sigma^{-1}D_AD_B\sigma \nonumber \\
&+\frac12g_{AB}\biggl\{2(n-2)\sigma^{-1}D^2\sigma-(n-2)(n-3)\sigma^{-2}[k-(D\sigma)^2]\biggl\}, \\
G_{ij}=&\gamma _{ij}\biggl\{-\frac12\sigma^2({}^{(2)}{R})+(n-3)\sigma D^2\sigma-\frac12(n-3)(n-4)[k-(D\sigma)^2]\biggl\},
\end{align}
where ${}^{(2)}{R}$ is the Ricci scalar on $M^2$ and $D^2\sigma:=g^{AB}D_AD_B\sigma$. (See Appendix A in~\cite{Maeda:2007uu} for derivation.)

In general relativity, the generalized Misner-Sharp quasi-local mass~\cite{Misner:1964je,Maeda:2006pm} is defined for the spacetime (\ref{eq:ansatz}) as
\begin{align}
\label{qlm}
m_{\rm MS}:= \frac{(n-2)V_{n-2}^{(k)}}{2\kappa_n}\sigma^{n-3}[k-(D \sigma)^2],
\end{align} 
where a constant $V_{n-2}^{(k)}$ denotes the volume of $K_{(k)}^{n-2}$ if it is compact.
Among the basic properties of $m_{\rm MS}$ studied in~\cite{Hayward:1994bu,Maeda:2007uu}, we will use the fact that $m_{\rm MS}$ converges to the ADM mass at spacelike infinity in an asymptotically flat spacetime.

\subsection{Exact solutions for $p=-(n-3)\rho/(n+1)$}

Now we derive the general solution for an equation of state $p=-(n-3)\rho/(n+1)$ by adopting the following comoving coordinates:
\begin{align}
\label{ansatz}
\begin{aligned}
&\D s^2=-\frac{\alpha(x)}{\beta(x)}\D t^2+\beta(x)^{-2(n-4)/(n-3)}\biggl(\frac{\beta(x)}{\alpha(x)}\D x^2+x^2\beta(x)^2\gamma_{ij}\D z^i\D z^j\biggl),\\
&u^\mu\frac{\partial}{\partial x^\mu}=\sqrt{\frac{\beta}{\alpha}}\frac{\partial}{\partial t}.
\end{aligned} 
\end{align} 
This is a static spacetime in the domain with $\alpha>0$ and $\beta>0$ or $\alpha<0$ and $\beta<0$.
However, since the powers of $\beta$ in the metric are integers for $n=4$ and $5$ and rational numbers for $n\ge 6$, the domain with $\alpha<0$ and $\beta<0$ are allowed only for $n=4$ and $5$.
The author has discovered the metric ansatz (\ref{ansatz}) by trial and error based on the experience of performing a complete classification of solutions in the Einstein-Maxwell system in~\cite{Maeda:2016ddh}.

With the metric (\ref{ansatz}), a combination $-(n-3){{\cal E}^t}_t-(n-5){{\cal E}^x}_x+2(n-2){{\cal E}^2}_2=0$ of the Einstein equations (\ref{EFE-0}) gives
\begin{align}
\kappa_n\{(n-3)\rho+(n+1)p\}=(n-2)x^{-1}\beta^{(n-5)/(n-3)}\left\{x \alpha''+(n-2)\alpha'\right\}.
\end{align} 
With an equation of state $(n-3)\rho+(n+1)p=0$, the above equation is integrated to give
\begin{align}
\alpha(x)=\alpha_0+\frac{\alpha_1}{x^{n-3}},\label{def-alpha}
\end{align} 
where $\alpha_0$ and $\alpha_1$ are integration constants.
Substituting Eq.~(\ref{def-alpha}) into a combination ${{\cal E}^x}_x-{{\cal E}^2}_2=0$, we obtain the following master equation for $\beta(x)$:
\begin{align}
0=&\biggl\{(n-4)\alpha_0+\frac{(n-2)\alpha_1}{x^{n-3}}\biggl\}x\beta'+\biggl(\alpha_0+\frac{\alpha_1}{x^{n-3}}\biggl)x^2\beta''+2(n-3)(k-\alpha_0\beta).\label{master-beta}
\end{align} 
The general solution to Eq.~(\ref{master-beta}) for $\alpha_0\ne 0$ is 
\begin{align}
\beta(x)=\frac{k}{\alpha_0}+\frac{\eta}{x^{n-3}}-\zeta x^2\biggl(\alpha_0+\frac{\alpha_1}{x^{n-3}}\biggl)^{(n-1)/(n-3)},
\end{align} 
where $\eta$ and $\zeta$ are integration constants.
The general solution to Eq.~(\ref{master-beta}) for $\alpha_0=0$ (and then $\alpha_1\ne 0$ is required) is 
\begin{align}
\beta(x)=\beta_0+\frac{\beta_1}{x^{n-3}}-\frac{k}{(n-3)\alpha_1}x^{n-3},
\end{align} 
where $\beta_0$ and $\beta_1$ are integration constants.
The signs of the functions $\alpha(x)$ and $\beta(x)$ are determined by the values of the parameters $\alpha_0$, $\alpha_1$, $\eta$, $\zeta$, $\beta_0$, and $\beta_1$, and we focus on spacetimes with a Lorentzian metric.

\subsubsection{Semiz class-I solution}

The metric functions in Eq.~(\ref{ansatz}) and the corresponding energy density $\rho$ of the general solution for $\alpha_0\ne 0$ are given by 
\begin{align}
\label{Semiz-I}
\begin{aligned}
&\alpha(x)=\alpha_0+\frac{\alpha_1}{x^{n-3}},\qquad \beta(x)=\frac{k}{\alpha_0}+\frac{\eta}{x^{n-3}}-\zeta x^2\biggl(\alpha_0+\frac{\alpha_1}{x^{n-3}}\biggl)^{(n-1)/(n-3)},\\
&\rho=-\frac{n+1}{n-3}p=\frac{(n^2-1)(n-2)\alpha_0^2\zeta}{2(n-3)\kappa_n}\frac{\alpha(x)^{2/(n-3)}}{\beta(x)^{2/(n-3)}}.
\end{aligned} 
\end{align} 
We refer to this solution as the Semiz class-I solution.
This solution for $n=4$ and $k=1$ corresponds to Semiz's solution for $C_1\ne 0$~\cite{Semiz:2020lxj} and its particular case with $\alpha_1=\alpha_0^2\eta$ was obtained by Ivanov by the Buchdahl transformation from the (anti-)de~Sitter solution~\cite{Ivanov:2001sg}. 
(See the vacuum dual (\ref{Semiz-I-dual}) to the Semiz class-I solution below.)

To see that Eq.~(\ref{Semiz-I}) is actually a two-parameter family of solutions, we perform coordinate transformations $(t,x)\to({\bar t},{\bar x})$ such that ${\bar t}=(n-3)|\alpha_0|t$ and ${\bar x}=(\alpha_0x^{n-3}+\alpha_1)/[(n-3)^2|\alpha_0|^2]$ together with redefinitions of the parameters ${\bar \eta}:=\eta-k\alpha_1/\alpha_0^2$ and ${\bar \zeta}:=\zeta (n-3)^{2(n-1)/(n-3)}|\alpha_0|^{2(n-1)/(n-3)}$.
Then the Semiz class-I solution is written as
\begin{align}
\label{Semiz-I-twopara}
\begin{aligned}
&\D s^2=-\frac{{\bar x}}{{\bar \beta}({\bar x})}\D {\bar t}^2+{\bar \beta}({\bar x})^{-2(n-4)/(n-3)}\biggl(\frac{{\bar \beta}({\bar x})}{{\bar x}}\D {\bar x}^2+{\bar \beta}({\bar x})^2\gamma_{ij}\D z^i\D z^j\biggl),\\
&\rho=-\frac{n+1}{n-3}p=\frac{(n^2-1)(n-2){\bar \zeta}}{2(n-3)^{3}\kappa_n}\frac{{\bar x}^{2/(n-3)}}{{\bar \beta}({\bar x})^{2/(n-3)}},\\
&{\bar \beta}({\bar x}):=x({\bar x})^{n-3}\beta(x({\bar x}))={\bar \eta}+k(n-3)^2{\bar x}-{\bar \zeta}{\bar x}^{(n-1)/(n-3)},
\end{aligned} 
\end{align} 
which is characterized by two parameters ${\bar \eta}$ and ${\bar \zeta}$.
The expression (\ref{Semiz-I-twopara}) shows that $\rho$ blows when ${\bar \beta}({\bar x})=0$ holds for ${\bar \zeta}\ne 0$.

For $k=\pm1$, the solution (\ref{Semiz-I-twopara}) in the vacuum limit ${\bar \zeta}\to 0$ is expressed by coordinate transformations ${\bar t}=k\tau/(n-3)$ and $r=[{\bar\eta}+k(n-3)^2{\bar x}]^{1/(n-3)}$ and a redefinition of the parameter $2M:=k{\bar \eta}$ as
\begin{align}
\label{ST}
\begin{aligned}
&\D s^2=-\biggl(k-\frac{2M}{r^{n-3}}\biggl)\D \tau^2+\biggl(k-\frac{2M}{r^{n-3}}\biggl)^{-1}\D r^2+r^2\gamma_{ij}\D z^i\D z^j.
\end{aligned} 
\end{align} 
This is the topological Schwarzschild-Tangherlini solution.
We note that ${\tilde M}:=(n-2)V_{n-2}^{(k)}M/\kappa _n$ coincides with the generalized Misner-Sharp mass (\ref{qlm}).
For $k=0$, the form (\ref{Semiz-I-twopara}) admits the following vacuum limit ${\bar \zeta}\to 0$:
\begin{align}
&\D s^2=-\frac{{\bar x}}{{\bar \eta}}\D {\bar t}^2+\frac{{\bar \eta}^{-(n-5)/(n-3)}}{{\bar x}}\D {\bar x}^2+{\bar \eta}^{2/(n-3)}\gamma_{ij}\D z^i\D z^j.
\end{align} 
This is the Ricci-flat direct-product solution, which is a direct product spacetime ${\rm M}_2\times K_{(0)}^{n-2}$ of a two-dimensional Minkowski spacetime ${\rm M}_2$ and an $(n-2)$-dimensional Ricci-flat space $K_{(0)}^{n-2}$.

Probably, the best expression of the Semiz class-I solution for $k=\pm 1$ is obtained from the metric (\ref{ansatz}) with Eq.~(\ref{Semiz-I}) by coordinate transformations $(t,x)\to({\tilde t},r)$ such that $t=(k/\alpha_0){\tilde t}$ and $x^{n-3}=(r^{n-3}-\eta)\alpha_0/k$ together with redefinitions of the parameters $2M:=k(\eta-k\alpha_1/\alpha_0^2)$ and ${\tilde\zeta}:=\zeta(\alpha_0^2/k^2)^{(n-1)/(n-3)}$.
Then the solution becomes
\begin{align}
\label{Semiz-I-twopara2}
\begin{aligned}
&\D s^2=-f(r)\D {\tilde t}^2+h(r)^{-2(n-4)/(n-3)}\biggl(f(r)^{-1}\D r^2+r^2h(r)^2\gamma_{ij}\D z^i\D z^j\biggl),\\
&\rho=-\frac{n+1}{n-3}p=\frac{(n^2-1)(n-2)k^2{\tilde\zeta}}{2(n-3)\kappa_n}f(r)^{2/(n-3)}
\end{aligned} 
\end{align} 
with
\begin{align}
&f(r)=\frac{kr^{n-3}-2M}{r^{n-3}-{\tilde\zeta}\left(kr^{n-3}-2M\right)^{(n-1)/(n-3)}},\\
&h(r)=1-\frac{{\tilde\zeta}}{r^{n-3}}\left(kr^{n-3}-2M\right)^{(n-1)/(n-3)}.
\end{align} 
Although the coordinate transformations are singular, this form of the solution is valid also for $k=0$.
The form (\ref{Semiz-I-twopara2}) clearly shows that the vacuum limit ${\tilde \zeta}\to 0$ is the topological Schwarzschild-Tangherlini solution (\ref{ST}) for any $k$.
The solution (\ref{Semiz-I-twopara2}) with $k=0$ is also given by Eq.~(\ref{ST}) with $k=0$, which is shown by further coordinate transformations.

\subsubsection{Semiz class-II solution}

The metric functions in Eq.~(\ref{ansatz}) and the corresponding energy density $\rho$ of the general solution for $\alpha_0=0$ (and then $\alpha_1\ne 0$) are given by 
\begin{align}
\label{Semiz-II}
\begin{aligned}
&\alpha(x)=\frac{\alpha_1}{x^{n-3}},\qquad \beta(x)=\beta_0+\frac{\beta_1}{x^{n-3}}-\frac{k}{(n-3)\alpha_1}x^{n-3},\\
&\rho=-\frac{n+1}{n-3}p=\frac{(n+1)(n-2)k}{2\kappa_nx^2\beta(x)^{2/(n-3)}}.
\end{aligned} 
\end{align} 
We refer to this solution as the Semiz class-II solution.
This solution for $n=4$ and $k=1$ corresponds to Semiz's solution for $C_1=0$~\cite{Semiz:2020lxj}.

For $\beta_0\ne 0$, by coordinate transformations $(t,x)\to({\bar t},r)$ such that ${\bar t}=\beta_0^{-1}t$ and $r^{n-3}=\beta_0x^{n-3}+\beta_1$ together with a redefinition of the parameter $2M=-\alpha_1\beta_0^2$, the solution is written as
\begin{align}
\label{Semiz-II-twopara}
\begin{aligned}
&\D s^2=\frac{2M}{r^{n-3}h(r)}\D {\bar t}^2-h(r)^{-2(n-4)/(n-3)}\biggl(\frac{r^{n-3}h(r)}{2M}\D r^2-r^2h(r)^2\gamma_{ij}\D z^i\D z^j\biggl),\\
&\rho=-\frac{n+1}{n-3}p=\frac{(n+1)(n-2)k}{2\kappa_nr^2h(r)^{2/(n-3)}},\\
&h(r):=x(r)^{n-3}\beta(x(r))/r^{n-3}=1+\frac{k(r^{n-3}-\beta_1)^2}{2(n-3)Mr^{n-3}},
\end{aligned} 
\end{align} 
which is characterized by two parameters $\beta_1$ and $M$.
The form (\ref{Semiz-II-twopara}) shows that the Semiz class-II solution with $\beta_0\ne 0$ in the vacuum case ($k=0$) is the topological Schwarzschild-Tangherlini solution (\ref{ST}) with $k=0$.

For $\beta_0=0$ with $\beta_1\ne 0$, by coordinate transformations $(t,x)\to({\bar t},r)$ such that
\begin{align}
{\bar t}=\sqrt{|\alpha_1|}|\beta_1|^{-(n-1)/[2(n-3)]}t,\qquad r=\frac{x^{n-3}}{(n-3)\sqrt{|\beta_1||\alpha_1|}},
\end{align} 
the Semiz class-II solution becomes
\begin{align}
\label{Semiz-II-twopara2}
\begin{aligned}
&\D s^2=|\beta_1|^{2/(n-3)}\biggl[-\frac{1}{{\bar\beta}(r)}\frac{\alpha_1}{|\alpha_1|}\D {\bar t}^2+{\bar\beta}(r)^{-(n-5)/(n-3)}\frac{|\alpha_1|}{\alpha_1}\D r^2+{\bar\beta}(r)^{2/(n-3)}\gamma_{ij}\D z^i\D z^j\biggl],\\
&{\bar\beta}(r):=\frac{x(r)^{n-3}\beta(x(r))}{|\beta_1|}=\frac{\beta_1}{|\beta_1|}-k(n-3)\frac{|\alpha_1|}{\alpha_1}r^2,\\
&\rho=-\frac{n+1}{n-3}p=\frac{(n+1)(n-2)k}{2\kappa_n|\beta_1|^{2/(n-3)}{\bar\beta}(r)^{2/(n-3)}}.
\end{aligned} 
\end{align} 
The form (\ref{Semiz-II-twopara2}) shows that the Semiz class-II solution with $\beta_0=0$ in the vacuum case ($k=0$) is the Ricci-flat direct-product solution ${\rm M}_2\times K_{(0)}^{n-2}$.

Lastly, for $\beta_0=\beta_1=0$ (and then $k\ne 0$ is required), by coordinate transformations $(t,x)\to({\bar t},r)$ such that
\begin{align}
&{\bar t}=\sqrt{\alpha_1}t,\qquad r=\biggl(-\frac{k}{(n-3)\alpha_1}\biggl)^{1/(n-3)}x^2,
\end{align} 
the Semiz class-II solution becomes
\begin{align}
\label{Semiz-II-last}
\begin{aligned}
&\D s^2=-\frac{1}{r^{n-3}}\D {\bar t}^2-\frac{n-3}{4k}\D r^2+r^2\gamma_{ij}\D z^i\D z^j,\\
&\rho=-\frac{n+1}{n-3}p=\frac{(n+1)(n-2)k}{2\kappa_nr^2}.
\end{aligned} 
\end{align} 
This solution does not admit a vacuum limit and requires $k=-1$ for the Lorentzian signature.
Actually, the solution (\ref{Semiz-II-last}) is a special case with $\chi=-(n-3)/(n+1)$ of the generalized Tolman-VI solution for an equation of state $p=\chi\rho$ given in Appendix~\ref{app:HW}.

\subsection{Semiz-class solutions as vacuum duals}

Here we show that the Semiz class solutions are duals to the $\Lambda$-vacuum solutions through the Buchdahl transformation.
In particular, the Semiz class-I solution is dual to the topological Schwarzschild-Tangherlini-(A)dS solution given by
\begin{align}
\label{ST-dS}
\begin{aligned}
\D s^2 =& -f(r)\D t^2+f(r)^{-1}\D r^2+r^2\gamma_{ij}\D z^i \D z^j,\\
f(r)=&k-\frac{2M}{r^{n-3}}-\frac{2\Lambda}{(n-1)(n-2)}r^2,
\end{aligned}
\end{align}
while the Semiz class-II solution is dual to the generalized Nariai solution, generalized anti-Nariai solution, and the Ricci-flat direct product solution for $k=1$, $k=-1$, and $k=0$, respectively.
The generalized Nariai (anti-Nariai) solution for $k=1$ with $\Lambda>0$ ($k=-1$ with $\Lambda<0$) is a direct-product solution ${\rm dS}_{2}\times K_{(1)}^{n-2}$ (${\rm AdS}_{2}\times K_{(-1)}^{n-2}$) of a two-dimensional de~Sitter (anti-de~Sitter) spacetime and an $(n-2)$-dimensional Einstein space with positive (negative) Ricci curvature, of which line-element may be written as
\begin{align}
\label{nariai}
\begin{aligned}
&\D s^2=-f(r)\D t^2+f(r)^{-1}\D r^2 +\frac{k(n-2)(n-3)}{2\Lambda}\gamma _{ij}\D z^i\D z^j,\\
&f(r)=f_0+f_1r-\frac{2\Lambda}{n-2}r^2,
\end{aligned}
\end{align}
where $f_0$ and $f_1$ are arbitrary constants.

\subsubsection{$n(\ge 4)$-dimensional Buchdahl transformation}

Now we present the $n(\ge 4)$-dimensional Buchdahl transformation which map a static solution to another in the system given by Eqs.~(\ref{EFE-0}) and (\ref{Tab-0})~\cite{Buchdahl:1956zz}.
(See also Sec.~10.11 in the textbook~\cite{Stephani:2003} for $n=4$.)
\begin{Prop}
\label{prop:Buchdahl}
Suppose that the following set
\begin{align}
\label{gauge-higher}
\begin{aligned}
&\D s^2=-\Omega(X)^{-2}\D t^2+\Omega(X)^{2/(n-3)}{\bar g}_{IJ}(X)\D X^I\D X^J,\\
&u^\mu\frac{\partial}{\partial x^\mu}=\Omega\frac{\partial}{\partial t},\qquad \rho=\rho_0(X),\qquad p=p_0(X)
\end{aligned}
\end{align}
is a solution to the Einstein equations (\ref{EFE-0}) with a perfect fluid (\ref{Tab-0}), where the indices $I$ and $J$ run from $1$ to $n-1$.
Then, the following set is also a solution:
\begin{align}
\label{p-rho-generated}
\begin{aligned}
&\D s^2=-\Omega(X)^{2}\D t^2+\Omega(X)^{-2/(n-3)}{\bar g}_{IJ}(X)\D X^I\D X^J,\\
&u^\mu\frac{\partial}{\partial x^\mu}=\Omega^{-1}\frac{\partial}{\partial t},\quad \rho=-\biggl\{\rho_0+\frac{2(n-1)}{n-3}p_0\biggl\}{\Omega}^{4/(n-3)},\quad p=p_0\Omega^{4/(n-3)}.
\end{aligned}
\end{align}
\end{Prop}
{\it Proof}. 
For the metric~(\ref{gauge-higher}), the Einstein tensor $G_{\mu\nu}$ is decomposed as
\begin{align}
{G}{}_{tt}=&\frac12\Omega^{-2(n-2)/(n-3)}\left\{{\bar R}-\frac{2(n-2)}{n-3}{\bar D}^2\ln\Omega-\frac{n-2}{n-3}({\bar D}\ln\Omega)^2\right\},\label{Ein-decomp3} \\
{G}{}_{IJ}=&{\bar G}_{IJ}-\frac{n-2}{n-3}({\bar D}_I\ln\Omega)({\bar D}_J\ln\Omega) +\frac{n-2}{2(n-3)}{\bar g}_{IJ}({\bar D}\ln\Omega)^2,\label{Ein-decomp4}
\end{align}
where ${\bar D}_I$ is the covariant derivative with respect to ${\bar g}_{IJ}$ and we have defined ${\bar D}^2:={\bar g}^{IJ}{\bar D}_I{\bar D}_J$ and $({\bar D}\ln\Omega)^2:={\bar g}^{IJ}({\bar D}_I\ln\Omega)({\bar D}_J\ln\Omega)$.
Here ${\bar R}_{IJ}$, ${\bar R}$, and ${\bar G}_{IJ}$ are the Ricci tensor, Ricci scalar, and Einstein tensor constructed from ${\bar g}_{IJ}$, respectively.
(See Appendix A in~\cite{Maeda:2019tqs} for derivation.)
Then, Eqs.~(\ref{EFE-0}) and (\ref{Tab-0}) for the set (\ref{gauge-higher}) reduce to the following set of equations:
\begin{align}
&(n-3){\bar R}_{IJ}=(n-2)({\bar D}_I\ln\Omega)({\bar D}_J\ln\Omega)-2\kappa_np\Omega^{2/(n-3)}{\bar g}_{IJ},\\
&(n-2){\bar D}^2\ln\Omega=-\kappa_n[(n-3)\rho+(n-1)p]\Omega^{2/(n-3)}.
\end{align}
The above field equations are invariant under the following transformations:
\begin{align}
\Omega={\tilde\Omega}^{-1},\quad p={\tilde p}{\tilde\Omega}^{4/(n-3)},\quad \rho=-\biggl\{{\tilde\rho}+\frac{2(n-1)}{n-3}{\tilde p}\biggl\}{\tilde\Omega}^{4/(n-3)}.
\end{align}
\qed

Proposition~\ref{prop:Buchdahl} is an $n(\ge 4)$-dimensional generalization of the solution-generating transformation presented in~\cite{Buchdahl:1956zz} for $n=4$.
However, since Buchdahl presented the expressions (\ref{Ein-decomp3}) and (\ref{Ein-decomp4}) for arbitrary $n(\ge 4)$ in a different notation\footnote{In~\cite{Buchdahl:1956zz}, Buchdahl used $P_{\mu\nu}$ for the Einstein tensor and $G_{\mu\nu}$ for the Ricci tensor.}, Proposition~\ref{prop:Buchdahl} should be attributed to him\footnote{The metric (\ref{gauge-higher}) for the most general static spacetime in $n(\ge 4)$ dimensions was introduced by Buchdahl in~\cite{Buchdahl1954}.}.
If a seed solution (\ref{gauge-higher}) satisfies $p_0=\chi\rho_0$, the generated solution (\ref{p-rho-generated}) satisfies $p={\tilde\chi}\rho$, where 
\begin{align}
{\tilde \chi}:=-\frac{(n-3)\chi}{(n-3)+2(n-1)\chi}.\label{chi-2}
\end{align}
In particular, $\chi={\tilde\chi}$ holds for $\chi(={\tilde\chi})=-(n-3)/(n-1)$, for which the strong energy condition is marginally satisfied. (See Eq.~(\ref{SEC-I}) below with $p_1=p_2\equiv p$.)
It was claimed in~\cite{Chernin:2001nu} that the general spherically symmmetric static solution had been obtained in this case with $n=4$ (hence $\chi=-1/3$). However, a complete proof does not seem to have been given.
Another interesting equation of state is $\chi=-(n-3)/[2(n-1)]$, with which the dual solution satisfies $\rho=0$.
In this case with $n=4$ (hence $\chi=-1/6$), Semiz constructed a particular static solution with spherical symmetry by the Buchdahl transformation~\cite{Semiz:2022iyh}.

\subsubsection{Vacuum duals to the Semiz-class solutions}

Here we adopt the Buchdahl transformation to the Semiz-class solutions.
The metric generated by the Buchdahl transformation from the seed metric (\ref{ansatz}) is given by 
\begin{align}
\D s^2=&-\frac{\beta(x)}{\alpha(x)}\D t^2+\alpha(x)^{2/(n-3)}\biggl(\frac{\D x^2}{\alpha(x)\beta(x)}+x^2\gamma_{ij}\D z^i\D z^j\biggl).\label{new-solution}
\end{align}
Equation~(\ref{chi-2}) shows ${\tilde\chi}=-1$ for $\chi=-(n-3)/(n+1)$.
Therefore, the Semiz-class solutions are dual to $\Lambda$-vacuum solutions.

Substituting $\alpha(x)$ and $\beta(x)$ of the Semiz-I class solution (\ref{Semiz-I}) ($\alpha_0\ne 0$ is required) to the metric (\ref{new-solution}), the corresponding Einstein tensor is computed to give
\begin{align}
{G^\mu}_\nu=-\frac12(n-1)(n-2)\zeta \alpha_0^2{\delta^\mu}_\nu,
\end{align} 
so that its dual is a $\Lambda$-vacuum solution with $\Lambda=(n-1)(n-2)\zeta \alpha_0^2/2$.
By coordinate transformation $t=\alpha_0{\bar t}$ and $r=(\alpha_0x^{n-3}+\alpha_1)^{1/(n-3)}$, the dual solution is written as
\begin{align}
\label{Semiz-I-dual}
\begin{aligned}
&\D s^2=-f(r)\D {\bar t}^2+f(r)^{-1}\D r^2+r^2\gamma_{ij}\D z^i\D z^j,\\
&f(r)=k-\frac{k\alpha_1-\alpha_0^2\eta}{r^{n-3}}-\alpha_0^2\zeta r^2,
\end{aligned}
\end{align}
which is the topological Schwarzschild-Tangherlini-(A)dS solution (\ref{ST-dS}) with $2M=k\alpha_1-\alpha_0^2\eta$ and $\Lambda=(n-1)(n-2)\alpha_0^2\zeta/2$.

Substituting $\alpha(x)$ and $\beta(x)$ of the Semiz-II class solution (\ref{Semiz-II}) ($\alpha_1\ne 0$ is required) to the metric (\ref{new-solution}), the corresponding Einstein tensor is computed to give
\begin{align}
{G^\mu}_\nu=-\frac{(n-2)(n-3)k}{2\alpha_1^{2/(n-3)}}{\delta^\mu}_\nu,
\end{align} 
so that its dual is a $\Lambda$-vacuum solution with $\Lambda=(n-2)(n-3)k/(2\alpha_1^{2/(n-3)})$.
By a coordinate transformation $r=\alpha_1^{-(n-4)/(n-3)}x^{n-3}/(n-3)$, the dual solution is written as
\begin{align}
\label{Semiz-II-dual}
\begin{aligned}
&\D s^2=-h(r)\D t^2+h(r)^{-1}\D r^2+\alpha_1^{2/(n-3)}\gamma_{ij}\D z^i\D z^j,\\
&h(r)=\frac{\beta_1}{\alpha_1}+\frac{(n-3)\beta_0}{\alpha_1^{1/(n-3)}}r-\frac{k(n-3)}{\alpha_1^{2/(n-3)}}r^2
\end{aligned}
\end{align}
which is the generalized Nariai (anti-Nariai) solution (\ref{nariai}) for $k=1$ ($k=-1$) with $\Lambda=(n-2)(n-3)k/(2\alpha_1^{2/(n-3)})$.
For $k=0$, Eq.~(\ref{Semiz-II-dual}) is the Ricci-flat direct-product solution ${\rm M}_2\times K_{(0)}^{n-2}$.

\section{Properties of the Semiz class-I solution with spherical symmetry}
\label{sec:Semiz-property}

In this section, we study the Semiz class-I solution (\ref{Semiz-I-twopara2}) with spherical symmetry, where $K_{(k)}^{n-2}$ is an $(n-2)$-dimensional sphere ${\rm S}^{n-2}$ (and hence $k=1$).
Here we present again the solution without tildes for simplicity;
\begin{align}
\label{Semiz-I-twopara-k=1}
\begin{aligned}
&\D s^2=-f(r)\D t^2+h(r)^{-2(n-4)/(n-3)}f(r)^{-1}\D r^2+r^2h(r)^{2/(n-3)}\gamma_{ij}\D z^i\D z^j,\\
&\rho=-\frac{n+1}{n-3}p=\frac{(n^2-1)(n-2){\zeta}}{2(n-3)\kappa_n}f(r)^{2/(n-3)},
\end{aligned} 
\end{align} 
where $f$ and $h$ are given by 
\begin{align}
&f(r)=\frac{r^{n-3}-2M}{r^{n-3}-{\zeta}\left(r^{n-3}-2M\right)^{(n-1)/(n-3)}},\label{def-f} \\
&h(r)=1-\frac{{\zeta}}{r^{n-3}}\left(r^{n-3}-2M\right)^{(n-1)/(n-3)}.\label{def-h} 
\end{align} 
With $\zeta=0$, the solution reduces to the Schwarzschild-Tangherlini vacuum solution and hereafter we assume $\zeta\ne 0$.

For convenience, we introduce a new coordinate $y:=r^{n-3}-2M$, with which the solution (\ref{Semiz-I-twopara-k=1}) is written as
\begin{align}
&\D s^2=-\frac{y}{\Pi(y)}\D t^2+\frac{\Pi(y)^{-(n-5)/(n-3)}}{(n-3)^2y}\D y^2+\Pi(y)^{2/(n-3)}\gamma_{ij}\D z^i\D z^j,\label{Semiz-I-y}\\
&\rho=-\frac{n+1}{n-3}p=\frac{(n^2-1)(n-2)\zeta}{2(n-3)\kappa_n}\biggl(\frac{y}{\Pi(y)}\biggl)^{2/(n-3)},\label{rho-y}\\
&\Pi(y):=2M+y-\zeta y^{(n-1)/(n-3)}.
\end{align} 
Hereafter we will study the solution in this coordinate system.
First of all, reality of the metric with the Lorentzian signature $(-,+,\cdots,+)$ restricts the domain of $y$.
In fact, $\Pi$ must be positive for $n\ge 5$ and, in addition, $y$ must be positive for $n\ge 6$.
By these constraints, $y\to \infty$ is allowed for $n=4$ with any $\zeta$ and $n\ge 5$ with $\zeta<0$, while $y\to -\infty$ is allowed for $n=4$ with any $\zeta$ and $n=5$ with $\zeta<0$.
Properties of the boundaries $\Pi=0$ and $y=0$ will be studied in the following subsections.

In the spherically symmetric case, we express the volume of ${\rm S}^{n-2}$ as $V_{n-2}^{(1)}={\cal A}_{n-2}$, where ${\cal A}_{n-2}$ is given in terms of the Gamma function $\Gamma(x)$ as
\begin{align}
{\cal A}_{n-2} :=\frac{2\pi^{(n-1)/2}}{\Gamma((n-1)/2)}.\label{unitarea}
\end{align}
Then, with the areal radius $\sigma=\Pi^{1/(n-3)}$, the Misner-Sharp mass (\ref{qlm}) is computed to give
\begin{align}
\label{qlm-Semiz-y}
m_{\rm MS}=&\frac{(n-2){\cal A}_{n-2}}{2\kappa_n}\Pi\left(1-y\Pi^{-1}{\Pi'}^2\right).
\end{align} 
In an asymptotically flat spacetime, $m_{\rm MS}$ converges to the ADM mass at spacelike infinity~\cite{Hayward:1994bu,Maeda:2007uu}.
Although $\lim_{y\to \pm \infty}R^{\mu\nu}_{\phantom{\mu}\phantom{\nu}\rho\sigma}=0$ is satisfied, Eq.~(\ref{qlm-Semiz-y}) blows up in the limit of $y\to \pm\infty$ as
\begin{align}
\lim_{y\to \pm \infty}m_{\rm MS} \simeq&-\frac{(n-1)^2(n-2){\cal A}_{n-2}\zeta^2}{2(n-3)^2\kappa_n}y^{(n+1)/(n-3)}.
\end{align} 
Therefore, the spacetime (\ref{Semiz-I-y}) with $\zeta\ne 0$ is asymptotically locally flat as $y\to \pm \infty$ if $y$ is a spacelike coordinate in the asymptotic regions.

\subsection{Scalar polynomial curvature singularity and metric reality}

\begin{figure}[htbp]
\begin{center}
\includegraphics[width=0.8\linewidth]{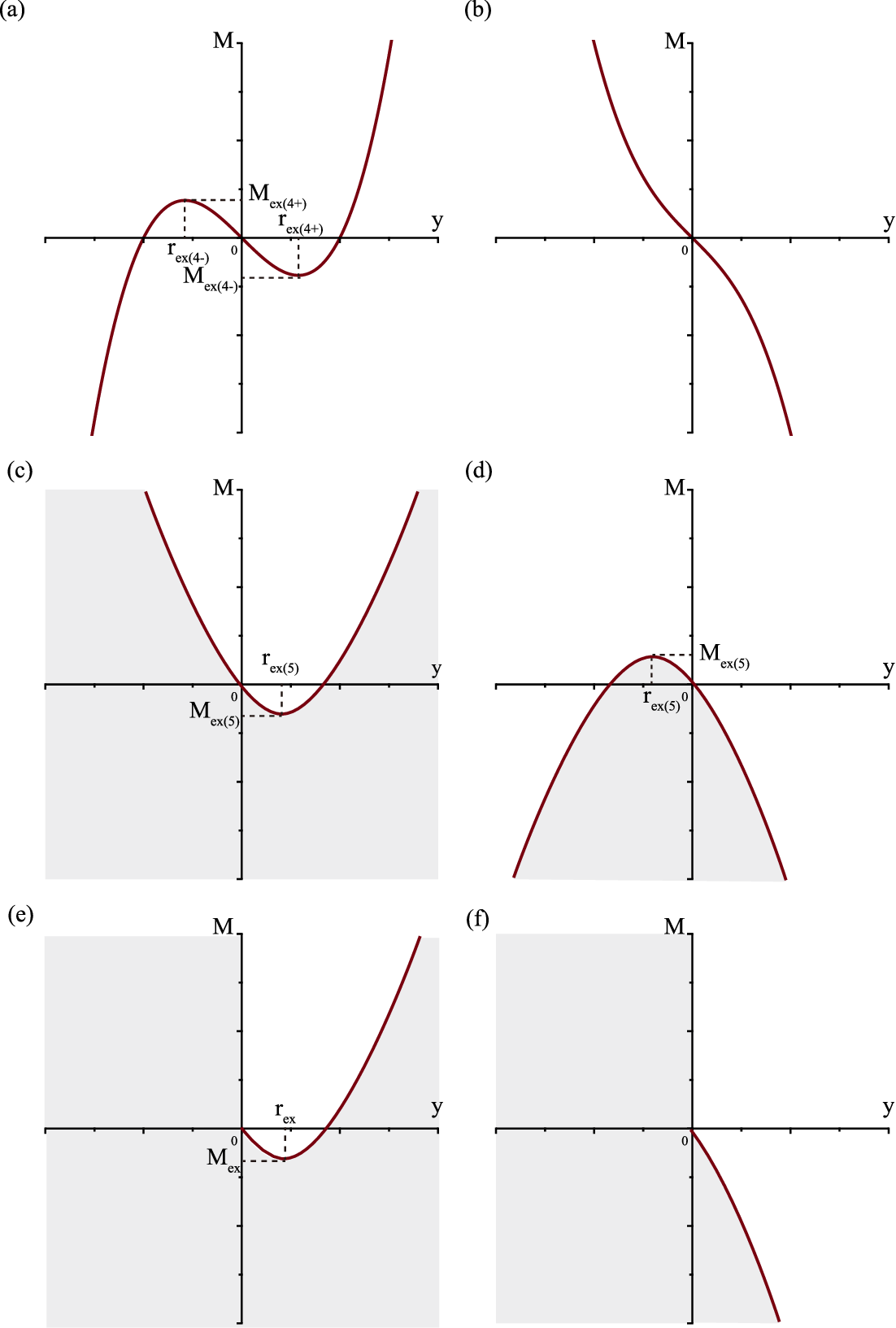}
\caption{\label{M-all} The forms of $M=M_{\rm s}(y)$ for (a) $n=4$ with $\zeta>0$, (b) $n=4$ with $\zeta<0$, (c) $n=5$ with $\zeta>0$, (d) $n=5$ with $\zeta<0$, (e) $n\ge 6$ with $\zeta>0$, and (f) $n\ge 6$ with $\zeta<0$. The metric (\ref{Semiz-I-y}) becomes complex and unphysical in shaded regions.}
\end{center}
\end{figure}

In the spacetime (\ref{Semiz-I-y}) with $M\ne 0$ (and $\zeta\ne 0$), the energy density $\rho$ blows up at $y=y_{\rm s}(\ne 0)$ determined by $\Pi(y_{\rm s})=0$ and therefore it is a scalar polynomial curvature singularity.
The relation between $M$ and $y_{\rm s}$ is given by 
\begin{align}
M=\frac12\left(-y_{\rm s}+\zeta y_{\rm s}^{(n-1)/(n-3)}\right)=:M_{\rm s}(y_{\rm s}).
\end{align} 
The constraint $\Pi(y)>0$ for $n\ge 5$ is equivalent to $M> M_{\rm s}(y)$.
The forms of $M=M_{\rm s}(y)$ depending on $n$ and the sign of $\zeta$ are drawn in Fig.~\ref{M-all}.
For $n=4$, there are two extrema at $y=\pm1/\sqrt{3\zeta}=:y_{{\rm ex}(4\pm)}$ for $\zeta>0$ and we have $M_{\rm s}(y_{4(\pm)})=\mp 1/(3\sqrt{3\zeta})=:M_{\rm ex(4\mp)}$, while $M_{\rm s}(y)$ is monotonically decreasing for $\zeta<0$.
For $n=5$, there is a single extremum at $y=1/(2\zeta)=:y_{{\rm ex}(5)}$ and we have $M_{\rm s}(y_{{\rm ex}(5)})=-1/(8\zeta)=:M_{\rm ex(5)}$.
For $n\ge 6$, there is a single local minimum at $y=[(n-3)/\{(n-1)\zeta\}]^{(n-3)/2}=:y_{\rm ex}$ for $\zeta>0$ and we have $M_{\rm s}(y_{\rm ex})=-y_{\rm ex}/(n-1)=:M_{\rm ex}$.

In the case of $M=0$, $\Pi(y)=0$ corresponds to $y=0$ or $1-\zeta y^{2/(n-3)}=0$ and the latter is a scalar polynomial curvature singularity.
In contrast, the spacetime is analytic at $y=0$ corresponding to the regular center $r=0$ for any $n(\ge 4)$.
This fact is obvious in the coordinate system (\ref{Semiz-I-twopara-k=1}) with $M=0$:
\begin{align}
&\D s^2=-\frac{\D t^2}{1-{\zeta}r^2}+(1-{\zeta}r^2)^{-(n-5)/(n-3)}\D r^2+r^2(1-{\zeta}r^2)^{2/(n-3)}\gamma_{ij}\D z^i\D z^j,\label{metric-M=0}\\
&\rho=\frac{(n^2-1)(n-2)\zeta}{2(n-3)\kappa_n(1-\zeta r^2)^{2/(n-3)}}.
\end{align}

\subsection{Parallelly propagated scalar curvature singularities}

Next we show that $y=0$ and $y\to +\infty$ are parallelly propagated (p.p.) scalar curvature singularities for $n\ge 6$, which are defined by the fact that some component of the Riemann tensor in a parallelly propagated frame blows up~\cite{Hawking:1973uf}.

For this purpose, we first derive components of the Riemann tensor $R_{(a)(b)(c)(d)}:=R_{\mu\nu\rho\sigma}E^\mu_{(a)}E^\nu_{(b)}E^\rho_{(c)}E^\sigma_{(d)}$ in a parallelly propagated orthonormal frame with basis vectors $E^\mu_{(a)}$, where $a=0,1,\cdots,n-1$, along an affinely parametrized and future-directed ingoing radial null geodesic $\gamma$ in the spacetime (\ref{Semiz-I-y}).
A tangent vector $k^\mu$ of $\gamma$ satisfying $k_\mu k^\mu=0$ and $k^\nu \nabla_\nu k^\mu=0$ is given by 
\begin{equation}
k^\mu\frac{\partial}{\partial x^\mu}=\frac{C}{\sqrt{2}}\biggl(\frac{\Pi}{y}\frac{\partial}{\partial t}-(n-3)\Pi^{(n-4)/(n-3)}\frac{\partial}{\partial y}\biggl),
\end{equation}
where the sign of a constant $C$ is chosen in such a way that $k^\mu$ is future-directed.
An orthogonal null vector to $k^\mu$ is given by 
\begin{equation}
l^\mu\frac{\partial}{\partial x^\mu}=\frac{1}{\sqrt{2}C}\biggl(\frac{\partial}{\partial t}+\frac{(n-3)y}{\Pi^{1/(n-3)}}\frac{\partial}{\partial y}\biggl),
\end{equation}
which satisfies $l_\mu l^\mu=0$ and $k_\mu l^\mu=-1$.
Then we consider vectors $E^\mu_{(i)}~(i=2,3,\cdots,n-1)$ on ${\rm S}^{n-2}$ given by
\begin{equation}
E^\mu_{(i)}\frac{\partial}{\partial x^\mu}=\frac{1}{\Pi^{1/(n-3)}}e^j_{(i)}\frac{\partial}{\partial z^j},
\end{equation}
which satisfy $g_{\mu\nu}E^\mu_{(i)}E^\nu_{(i)}=\delta_{(i)(j)}$, where $e^j_{(i)}$ are basis vectors on ${\rm S}^{n-2}$ satisfying 
\begin{align}
\gamma_{ij}e_{(k)}^{i}e_{(l)}^{j}=\delta_{(k)(l)}~\Leftrightarrow~\gamma^{ij}=\delta^{(k)(l)}e_{(k)}^{i}e_{(l)}^{j}.\label{e-K}
\end{align}
Since $k^\nu \nabla_\nu l^\mu=0$ and $k^\nu \nabla_\nu E_{(i)}^\mu=0$ are satisfied, we identify $E^\mu_{(0)}\equiv k^\mu$ and $E^\mu_{(1)}\equiv l^\mu$ and then $E^\mu_{(a)}=\{k^\mu, l^\mu,E^\mu_{(i)}\}$ are basis vectors in a parallelly propagated orthonormal frame along $\gamma$.
With the following expressions;
\begin{align}
{R}_{tyty}=&\frac{-(n-3)y\Pi\Pi''+(n-2)\Pi'(y\Pi'-\Pi)}{2(n-3)\Pi^3},\\ 
{R}_{titj}=&-\frac{(n-3)y\Pi'(y\Pi'-\Pi)}{2\Pi^2}\gamma_{ij},\\ 
{R}_{yiyj}=&-\frac{2y\Pi\Pi''-\Pi'(y\Pi'-\Pi)}{2(n-3)y\Pi^{2(n-4)/(n-3)}}\gamma_{ij},\\ 
{R}_{ijkl}=&\Pi^{2/(n-3)}\left(k-y\Pi^{-1}{\Pi'}^2\right)(\gamma_{ik}\gamma _{jl}-\gamma_{il}\gamma _{jk}),
\end{align}
non-zero components of $R_{(a)(b)(c)(d)}$ are computed to give
\begin{align}
R_{(0)(1)(0)(1)}=&-\frac{n-3}{2\Pi^{(n-1)/(n-3)}}\biggl\{(n-3)y\Pi\Pi''-(n-2)\Pi'(y\Pi'-\Pi)\biggl\},\\
R_{(0)(i)(0)(j)}=&-\frac12(n-3)C^2\Pi^{(n-5)/(n-3)}\Pi''\delta_{(i)(j)}, \\
R_{(0)(i)(1)(j)}=&R_{(1)(i)(0)(j)}=\frac{n-3}{2\Pi^{(n-1)/(n-3)}}\left\{y\Pi\Pi''-\Pi'(y\Pi'-\Pi)\right\}\delta_{(i)(j)}, \\
R_{(1)(i)(1)(j)}=&-\frac{(n-3)y^2\Pi''}{2C^2\Pi^{(n-1)/(n-3)}}\delta_{(i)(j)}, \\
R_{(i)(j)(k)(l)}=&\frac{1-y\Pi^{-1}{\Pi'}^2}{\Pi^{2/(n-3)}}(\delta_{(i)(k)}\delta_{(j)(l)}-\delta_{(i)(l)}\delta_{(j)(k)}).
\end{align}

By the expression $\Pi''(y)=-2(n-1)\zeta y^{-(n-5)/(n-3)}/(n-3)^2$ and the finiteness of $\Pi(0)$ and $\Pi'(0)$, $R_{(0)(i)(0)(j)}$ diverges as $y\to 0$ for $n\ge 6$, which shows that $y=0$ is a p.p. curvature singularity.
Also, since $\lim_{y\to \infty}\Pi\simeq -\zeta y^{(n-1)/(n-3)}$ holds, we obtain $\lim_{y\to\infty}R_{(0)(i)(0)(j)}\propto y^{2(n-5)/(n-3)^2}$, which shows that $y\to\infty$ is a p.p. curvature singularity for $n\ge 6$, too.
In contrast, $R_{(a)(b)(c)(d)}$ are finite as $y\to \pm\infty$ for $n=4$ and $5$.

\subsection{Causal nature of boundaries}

Here we clarify the causal nature of $y=0$, $y\to \pm\infty$, and $y=y_{\rm s}$ in the Penrose diagram.
Causal nature of a hypersurface with constant $y(=y_1)$ is determined by the two-dimensional Lorentzian portion with constant $z^i$ in the spacetime (\ref{Semiz-I-y}), of which conformally flat form is given by 
\begin{align}
&\D s_2^2=\frac{y}{\Pi(y)}(-\D t^2+\D y_*^2),\\
&y_*:=\int^y \frac{\Pi({\bar y})^{1/(n-3)}}{(n-3){\bar y}}\D {\bar y}.\label{def-y*}
\end{align} 
$y=y_1$ is non-null (null) if $y_*$ is finite (diverges) in the limit of $y\to y_1$.

Near $y=0$, we have 
\begin{align}
\label{Pi-y=0}
\Pi(y)\simeq \left\{
\begin{array}{ll}
2M & ({\rm for}~M\ne 0)\\
y & ({\rm for}~M= 0)
\end{array}
\right.,
\end{align}
which shows that $\lim_{y\to 0}y_*$ blows up for $M\ne 0$ and it is finite for $M= 0$.
Hence, $y=0$ is null and timelike for $M\ne 0$ and $M=0$, respectively.
Next, as $y\to\pm\infty$, we obtain $\lim_{y\to \pm \infty}y_*\propto y^{(n-1)/(n-3)^2}$, which blows up for any $n(\ge 4)$.
Therefore, $y\to \pm \infty$ are null.
Lastly, near $y=y_{\rm s}(\ne 0)$, we have $\Pi(y_{\rm s})\simeq \Pi_1(y-y_{\rm s})^b$, where $\Pi_1$ is a non-zero constant and $b$ is given by
\begin{align}
\label{Pi-y=ys-b}
b= \left\{
\begin{array}{ll}
2 & ({\rm at~an~extremum~of}~M=M_{\rm s}(y))\\
1 & ({\rm otherwise})
\end{array}
\right..
\end{align}
Since $y_*$ is finite as $y\to y_{\rm s}$ both for $b=1$ and $2$, the singularity at $y=y_{\rm s}$ is non-null.

Let us also check extendibility of these boundaries.
For this purpose, we need to study affinely parametrized radial null geodesics.
Since the spacetime (\ref{Semiz-I-y}) admits a hypersurface-orthogonal Killing vector $\xi^\mu\partial/\partial x^\mu=\partial/\partial t$, there is a conserved quantity $E:=-\xi_\mu k^\mu$ along a geodesic with its tangent vector $k^\mu$.
A future-directed radial null geodesic $\gamma$ is described by $x^\mu=(t(\lambda), y(\lambda),0,\cdots,0)$, where $\lambda$ is an affine parameter along $\gamma$.
Then, by the expression $E=y\Pi^{-1}(\D t/\D\lambda)$ and the null condition $\D s^2=0$, we obtain
\begin{equation}
\frac{\D y}{\D\lambda}=\pm (n-3)|E|\Pi^{(n-4)/(n-3)},
\end{equation}
which is integrated to give
\begin{equation}
\pm (n-3)|E|(\lambda-\lambda_0)=\int^y \Pi({\bar y})^{-(n-4)/(n-3)}\D {\bar y},\label{boundary}
\end{equation}
where $\lambda_0$ is an integration constant and the plus (minus) sign corresponds to outgoing (ingoing) $\gamma$.
Then, $y=y_1$ is null infinity if $\lambda$ diverges as $y\to y_1$.
If $y\to y_1$ is regular and corresponds to a finite value of $\lambda$, the spacetime is extendible beyond $y=y_1$.

First, $y=0$ is not null infinity because the expression (\ref{Pi-y=0}) shows that $\lambda$ is finite there.
Next, as $y\to\pm\infty$, the right-hand side of Eq.~(\ref{boundary}) behaves as
\begin{align}
\int^y \Pi({\bar y})^{-(n-4)/(n-3)}\D {\bar y}\simeq (-\zeta)^{-(n-4)/(n-3)}\int^y {\bar y}^{-(n-1)(n-4)/(n-3)^2}\D {\bar y}.\label{extend-int}
\end{align} 
Since Eq.~(\ref{extend-int}) diverges for $n=4$ and $5$ and remains finite for $n\ge 6$, $y\to \pm \infty$ are null infinities for $n=4$ and $5$, while a p.p. curvature singularity $y\to +\infty$ for $n\ge 6$ is not null infinity.
Lastly, near $y=y_{\rm s}$, we have $\Pi(y_{\rm s})\simeq \Pi_1(y-y_{\rm s})^b$ with $b$ given by Eq.~(\ref{Pi-y=ys-b}).
It shows that $\lambda$ diverges as $y\to y_{\rm s}$ only for $b=2$ with $n\ge 5$.
Therefore, the singularity $y=y_{\rm s}$ is not null infinity except for the case where $y=y_{\rm s}$ is an extremum of $M=M_{\rm s}(y)$ for $n\ge 5$.

Based on these results, one can draw the Penrose diagrams of the Semiz class-I solution with $k=1$ depending on $M$, $\zeta$, and $n$, as summarized in Tables~\ref{table:Penrose-n=4}, \ref{table:Penrose-n=5}, and~\ref{table:Penrose-n=6} for $n=4$, $n=5$, and $n\ge 6$, respectively.
\begin{table*}[htb]
\begin{center}
\caption{Penrose diagrams of the Semiz class-I solution with $k=1$ and $n=4$. If $M=M_{\rm s}(y)$ admits more than one real solution for a given $M$, they are represented as $y=y_{\rm s(i)}~(i=1,2,\cdots)$ satisfying $y_{\rm s(i)}<y_{\rm s(i+1)}$.}
\label{table:Penrose-n=4}
\scalebox{0.885}{
\begin{tabular}{|c|c|c|c|c|}\hline
$\zeta$ & $M$ & Domain of $y$ & Diagram in Fig.~\ref{Fig:Semiz-Penrose} \\ \hline
$\zeta>0$ & $M>M_{\rm ex(4+)}$ & $y_{\rm s}<y<\infty$ & C1 \\ \cline{3-4}
& & $-\infty<y<y_{\rm s}$ & NS7 \\ \cline{2-4}
& $M=M_{\rm ex(4+)}$ & $y_{\rm s(2)}<y<\infty$ & C1 \\ \cline{3-4}
& (then $y_{\rm s(1)}=y_{\rm ex(4-)}$) & $y_{\rm ex(4-)}<y<y_{\rm s(2)}$ & NS6 \\ \cline{3-4}
& & $-\infty<y<y_{\rm ex(4-)}$ & C1 \\ \cline{2-4}
& $0<M<M_{\rm ex(4+)}$ & $y_{\rm s(3)}<y<\infty$ & C1 \\ \cline{3-4}
& & $y_{\rm s(2)}<y<y_{\rm s(3)}$ & NS6 \\ \cline{3-4}
& & $y_{\rm s(1)}<y<y_{\rm s(2)}$ & NS5 \\ \cline{3-4}
& & $-\infty<y<y_{\rm s(1)}$ & C1 \\ \cline{2-4}
& $M=0$ & $y_{\rm s(3)}<y<\infty$ & C1 \\ \cline{3-4}
& (then $y_{\rm s(2)}=0$) & $0<y<y_{\rm s(3)}$ & NS4 \\ \cline{3-4}
& & $y_{\rm s(1)}<y<0$ & NS4 \\ \cline{3-4}
& & $-\infty<y<y_{\rm s(1)}$ & C1 \\ \cline{2-4}
& $M_{\rm ex(4-)}<M<0$ & $y_{\rm s(3)}<y<\infty$ & C1 \\ \cline{3-4}
& & $y_{\rm s(2)}<y<y_{\rm s(3)}$ & NS5 \\ \cline{3-4}
& & $y_{\rm s(1)}<y<y_{\rm s(2)}$ & NS6 \\ \cline{3-4}
& & $-\infty<y<y_{\rm s(1)}$ & C1 \\ \cline{2-4}
& $M=M_{\rm ex(4-)}$ & $y_{\rm ex(4+)}<y<\infty$ & C1 \\ \cline{3-4}
& (then $y_{\rm s(2)}=y_{\rm ex(4+)}$)& $y_{\rm s(1)}<y<y_{\rm ex(4+)}$ & NS6 \\ \cline{3-4}
& & $-\infty<y<y_{\rm s(1)}$ & C1 \\ \cline{2-4}
& $M<M_{\rm ex(4-)}$ & $y_{\rm s}<y<\infty$ & NS7 \\ \cline{3-4}
& & $-\infty<y<y_{\rm s}$ & C1 \\ \hline
$\zeta<0$ & $M>0$ & $y_{\rm s}<y<\infty$ & BH1 \\ \cline{3-4}
& & $-\infty<y<y_{\rm s}$ & NS1 \\ \cline{2-4}
& $M=0$ & $0\le y<\infty$ & R \\ \cline{3-4}
& & $-\infty<y\le 0$ & R \\ \cline{2-4}
& $M<0$ & $y_{\rm s}<y<\infty$ & NS1 \\ \cline{3-4}
& & $-\infty<y<y_{\rm s}$ & BH1 \\ \hline
\end{tabular}
}
\end{center}
\end{table*}

\begin{table*}[htb]
\begin{center}
\caption{Penrose diagrams of the Semiz class-I solution with $k=1$ and $n=5$. See the caption of Table~\ref{table:Penrose-n=4}.}
\label{table:Penrose-n=5}
\begin{tabular}{|c|c|c|c|c|}\hline
$\zeta$ & $M$ & Domain of $y$ & Diagram in Fig.~\ref{Fig:Semiz-Penrose} \\ \hline
$\zeta>0$ & $M>0$ & $y_{\rm s(1)}<y<y_{\rm s(2)}$ & NS6 \\ \cline{2-4}
& $M=0$ (then $y_{\rm s(1)}=0$) & $0\le y<y_{\rm s(2)}$ & NS4 \\ \cline{2-4}
& $M_{\rm ex(5)}<M<0$ & $y_{\rm s(1)}<y<y_{\rm s(2)}$ & NS5 \\ \hline
$\zeta<0$ & $M>M_{\rm ex(5)}$ & $-\infty<y<\infty$ & BH3 \\ \cline{2-4}
& $M=M_{\rm ex(5)}$ & $r_{\rm ex(5)}<y<\infty$ & BH2 \\ \cline{3-4}
& & $-\infty<y<r_{\rm ex(5)}$ & C2 \\ \cline{2-4}
& $0<M< M_{\rm ex(5)}$ & $y_{\rm s(2)}<y<\infty$ & BH1 \\ \cline{3-4}
& & $-\infty<y<y_{\rm s(1)}$ & C1 \\ \cline{2-4}
& $M=0$ & $0\le y<\infty$ & R \\ \cline{3-4}
& (then $y_{\rm s(2)}=0$) & $-\infty<y<y_{\rm s(1)}$ & C1 \\ \cline{2-4}
& $M<0$ & $y_{\rm s(2)}<y<\infty$ & NS1 \\ \cline{3-4}
& & $-\infty<y<y_{\rm s(1)}$ & C1 \\ \hline
\end{tabular}
\end{center}
\end{table*}

\begin{table*}[htb]
\begin{center}
\caption{Penrose diagrams of the Semiz class-I solution with $k=1$ and $n\ge 6$. See the caption of Table~\ref{table:Penrose-n=4}.}
\label{table:Penrose-n=6}
\begin{tabular}{|c|c|c|c|c|}\hline
$\zeta$ & $M$ & Domain of $y$ & Diagram in Fig.~\ref{Fig:Semiz-Penrose} \\ \hline
$\zeta>0$ & $M>0$ & $0<y<y_{\rm s}$ & NS8 \\ \cline{2-4}
& $M=0$ (then $y_{\rm s(1)}=0$) & $0\le y<y_{\rm s(2)}$ & NS4 \\ \cline{2-4}
& $M_{\rm ex}<M<0$ & $y_{\rm s(1)}<y<y_{\rm s(2)}$ & NS5 \\ \hline
$\zeta<0$ & $M>0$ & $0<y<\infty$ & NS9 \\ \cline{2-4}
& $M=0$ & $0\le y<\infty$ & NS3 \\ \cline{2-4}
& $M<0$ & $y_{\rm s}<y<\infty$ & NS2 \\ \hline
\end{tabular}
\end{center}
\end{table*}

\begin{figure}[htbp]
\begin{center}
\includegraphics[width=0.65\linewidth]{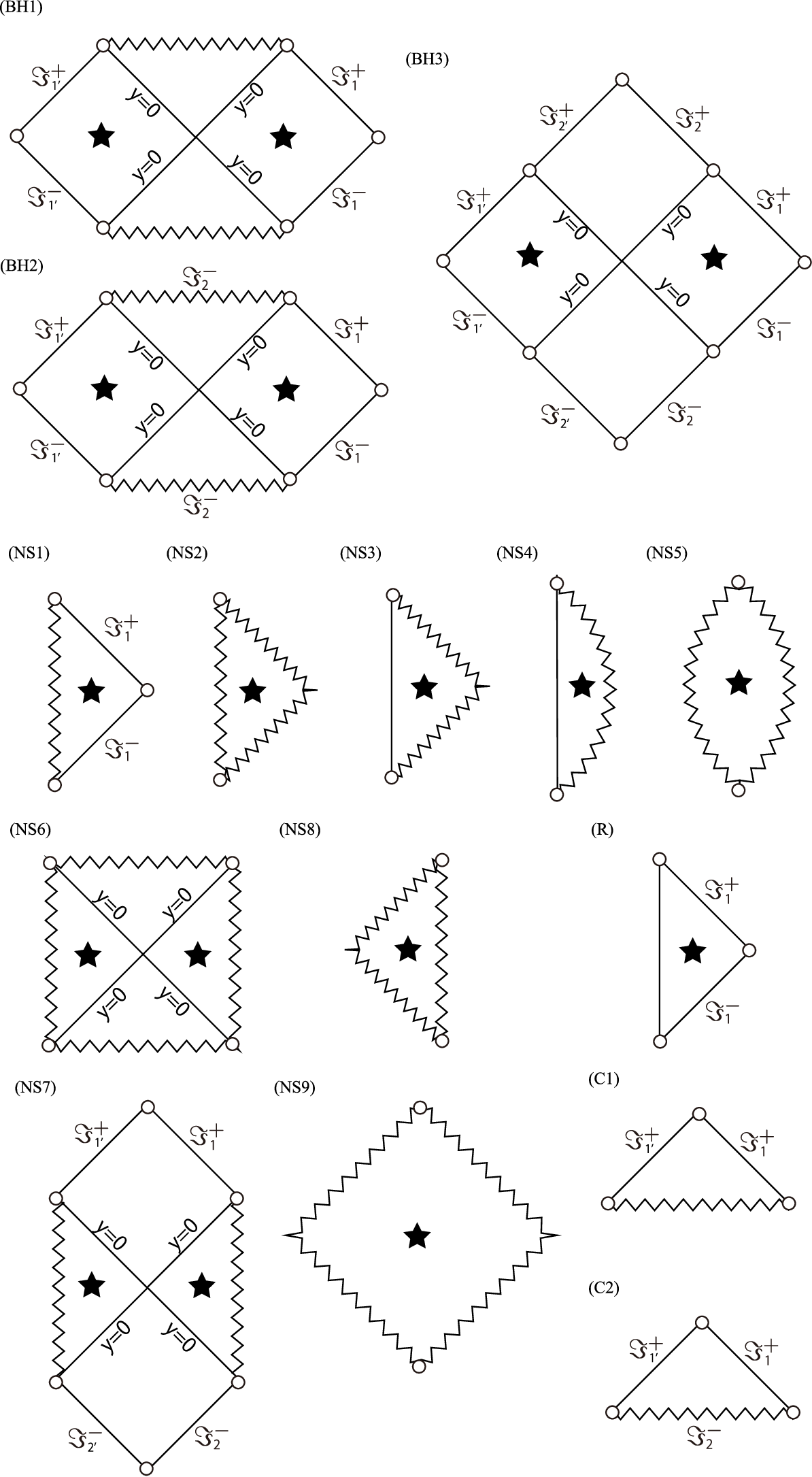}
\caption{\label{Fig:Semiz-Penrose} Possible Penrose diagrams of the Semiz class-I solution with $k=1$ and $\zeta\ne 0$.
A zig-zag line or curve is a curvature singularity and the spacetime is static in a region with a black star.
$\Im^{+(-)}$ stands for a future (past) null infinity and its indices are to distinguish different infinities.
BH, NS, R, and C stand for ``Black Hole'', ``Naked Singularity'', ``Regular'', and ``Cosmological'', respectively. Time-reversal (upside-down) diagrams are also possible for (C1) and (C2).}
\end{center}
\end{figure}

\subsection{Regular Killing horizon for $n=4$ and $5$}

The spacetime (\ref{Semiz-I-y}) admits a hypersurface orthogonal Killing vector $\xi^\mu(\partial/\partial x^\mu)=\partial/\partial t$, of which squared norm is $\xi_\mu\xi^\mu=-y/\Pi(y)$, so that $\xi^\mu$ is timelike (spacelike) in a region where $y/\Pi>(<)0$ holds.
A Killing horizon associated with $\xi^\mu$ is a regular null hypersurface $y=0$ for $M\ne 0$.
For $M=0$, we have $\xi_\mu\xi^\mu=-(1-\zeta y^{2/(n-3)})^{-1}$, so that $y=0$ is not a Killing horizon but a regular center $r=0$ as shown in the metric (\ref{metric-M=0}).
Therefore, we will focus on the case with $M\ne 0$ (and $\zeta\ne 0$).

In terms of the null coordinate defined by 
\begin{align}
v:=t+\int\frac{\Pi(y)^{1/(n-3)}}{(n-3)y}\D y,
\end{align} 
the metric (\ref{Semiz-I-y}) becomes
\begin{align}
&\D s^2=-\frac{y}{\Pi(y)}\D v^2+\frac{2}{(n-3)\Pi(y)^{(n-4)/(n-3)}}\D v\D y+\Pi(y)^{2/(n-3)}\gamma_{ij}\D z^i\D z^j.\label{Semiz-I-v}
\end{align} 
In terms of the coordinate $r$, this metric becomes 
\begin{align}
&\D s^2=-f(r)\D v^2+\frac{2}{h(r)^{(n-4)/(n-3)}}\D v\D r+r^2h(r)^{2/(n-3)}\gamma_{ij}\D z^i\D z^j
\end{align} 
where $f(r)$ and $h(r)$ are defined by Eqs.~(\ref{def-f}) and (\ref{def-h}), respectively.
Both of these metrics and their inverses are $C^\infty$ at $y=0$ and $r^{n-3}=2M$ for $n=4$ and $5$ and hence it is a Killing horizon.
In contrast, they are $C^1$ at a p.p. curvature singularity $y=0$ for $n\ge 6$.
In this subsection, we study the properties of $y=0$ in more detail.

\subsubsection{Matter field on and off the Killing horizon}

Here we study the matter field of the Semiz class-I solution (\ref{Semiz-I-y}) and discuss the standard energy conditions.
The standard energy conditions consist of the {\it null} energy condition (NEC), {\it weak} energy condition (WEC), {\it dominant} energy condition (DEC), and {\it strong} energy condition (SEC). (See section~2 in~\cite{Maeda:2018hqu}.)
It is emphasized that, although the matter field in a static region is a perfect fluid (\ref{Tab-0}), it is not the case in different regions.

In fact, in the region of $y\ne 0$, the matter field is described by the following anisotropic fluid in general;
\begin{align}
&{T}_{\mu\nu}=(\mu+p_2)u_\mu u_\nu+(p_1-p_2)s_\mu s_\nu +p_2g_{\mu\nu},\label{Tab-a}
\end{align}
where $\mu$, $p_1$, and $p_2$ are energy density, radial pressure, and tangential pressure of the fluid, respectively, and $u_\mu u^\mu=-1$, $s_\mu s^\mu=1$, and $u_\mu s^\mu=0$ hold.
This matter field becomes a perfect fluid if $p_1=p_2\equiv p$ is satisfied.
Equivalent expressions of the standard energy conditions for an anisotropic fluid (\ref{Tab-a}) are given by
\begin{align}
\mbox{NEC}:&~~\mu+p_i\ge 0~(i=1,2),\label{NEC-I}\\
\mbox{WEC}:&~~\mu\ge 0\mbox{~in addition to NEC},\label{WEC-I}\\
\mbox{DEC}:&~~\mu-p_i\ge 0~(i=1,2)\mbox{~in addition to WEC},\label{DEC-I}\\
\mbox{SEC}:&~~(n-3)\mu+p_1+(n-2)p_2\ge 0\mbox{~in addition to NEC}.\label{SEC-I}
\end{align}
(See section~3.1 in~\cite{Maeda:2018hqu}.)

Non-zero components of the energy-momentum tensor for the Semiz class-I solution (\ref{Semiz-I-y}) are 
\begin{align}
T^{t}_{~t}=-\rho, \qquad T^{y}_{~y}=-\frac{n-3}{n+1}\rho,\qquad T^{i}_{~j}=-\frac{n-3}{n+1}\rho\delta^{i}_{~j},
\end{align} 
where $\rho$ is given by Eq.~(\ref{rho-y}).
In a static region given by $y\Pi>0$ for $n=4$ and $y>0$ with $\Pi>0$ for $n\ge 5$, where $y$ is a spacelike coordinate, this matter field is described by Eq.~(\ref{Tab-a}) with 
\begin{align}
&\mu=\rho,\qquad p_1=p_2=-\frac{n-3}{n+1}\rho,\\
&u^\mu\frac{\partial}{\partial x^\mu}=\sqrt{\frac{\Pi}{y}}\frac{\partial}{\partial t},
\end{align}
which is certainly a perfect fluid (\ref{Tab-0}) obeying $p=-(n-3)\rho/(n+1)$.
In this case, all the standard energy conditions are satisfied (violated) for $\zeta>(<)0$.

On the other hand, in a dynamical region given by $y\Pi<0$ for $n=4$ and $y<0$ with $\Pi>0$ for $n\ge 5$, where $y$ is a timelike coordinate, the matter field is an anisotropic fluid (\ref{Tab-a}) with
\begin{align}
&\mu=\frac{n-3}{n+1}\rho,\qquad p_1=-\rho,\qquad p_2=-\frac{n-3}{n+1}\rho,\\
&u^\mu\frac{\partial}{\partial x^\mu}=\sqrt{-\frac{(n-3)^2y}{\Pi^{-(n-5)/(n-3)}}}\frac{\partial}{\partial y}, \qquad s^\mu\frac{\partial}{\partial x^\mu}=\sqrt{-\frac{\Pi}{y}}\frac{\partial}{\partial t}.
\end{align}
By the following expressions
\begin{align}
&\mu+p_1=-\frac{4}{n+1}\rho,\quad \mu+p_2=0,\\
&\mu-p_1=\frac{2(n-1)}{n+1}\rho,\quad \mu-p_2=\frac{2(n-3)}{n+1}\rho,\\
&(n-3)\mu+p_1+(n-2)p_2=-\frac{2(n-1)}{n+1}\rho
\end{align} 
and Eqs~(\ref{NEC-I})--(\ref{SEC-I}), all the standard energy conditions are violated for $\zeta>0$, while the NEC and SEC are satisfied for $\zeta<0$.

In contrast, a matter field on the Killing horizon $y=0$ for $n=4$ and $5$ is very non-trivial because $y=0$ is a coordinate singularity in the metric (\ref{Semiz-I-y}).
We identify the matter field on the Killing horizon $y=0$ with the regular metric (\ref{Semiz-I-v}) at $y=0$.
Non-zero components of the Einstein tensor in this coordinate system are given by
\begin{align}
G^{v}_{~v}=&-\frac{(n-1)(n+1)(n-2)}{2(n-3)}\zeta y^{2/(n-3)}\Pi^{-2/(n-3)}, \\
G^{v}_{~y}=&\frac{2(n-1)(n-2)}{(n-3)^2}\zeta y^{-(n-5)/(n-3)}\Pi^{-1/(n-3)}, \\
G^{y}_{~v}=&0,\\
G^{y}_{~y}=&-\frac12(n-1)(n-2)\zeta y^{2/(n-3)}\Pi^{-2/(n-3)}, \\
G^{i}_{~j}=&-\frac12(n-1)(n-2)\zeta y^{2/(n-3)}\Pi^{-2/(n-3)}\delta^{i}_{~j}.
\end{align} 
For $M\ne 0$, we obtain
\begin{align}
G^{v}_{~v}|_{y=0}=&G^{y}_{~v}|_{y=0}=G^{y}_{~y}|_{y=0}=G^{i}_{~j}|_{y=0}=0, \\
G^{v}_{~y}|_{y=0}=&\left\{
\begin{array}{ll}
0 & (n=4)\\
24\zeta/\sqrt{2M} & (n= 5)\\
\infty & (n\ge 6)\\
\end{array}
\right..
\end{align} 
Therefore, a matter field is absent on the Killing horizon for $n=4$, as shown in~\cite{Maeda:2021ukk}.
In contrast, divergence of $\lim_{y\to 0}G^{v}_{~y}$ for $n\ge 6$ suggests that $y=0$ is a p.p. curvature singularity.

For $n=5$, the matter field at $y=0$ is a null dust fluid, of which energy-momentum tensor is given by 
\begin{align}
&T_{\mu\nu}|_{y=0}=\Omega l_\mu l_\nu,\qquad l^\mu\frac{\partial}{\partial x^\mu}=l^0\frac{\partial}{\partial v},\label{typeII-null}
\end{align}
where $\Omega$ is the energy density and $l^\mu$ is a null vector satisfying $\Omega(l^0)^2=48\zeta/\kappa_5$.
This null dust fluid satisfies (violates) all the standard energy conditions for $\zeta>(<)0$.
(See section~4.2 in~\cite{Maeda:2018hqu}.)
If we assume that the null dust moves geodesically and is parametrized by an affine parameter $\lambda$, $\Omega$ and $l^0$ are given by 
\begin{align}
&\Omega=\frac{24\zeta}{\kappa_5M}(\lambda-\lambda_0)^2,\qquad l^0=\frac{\sqrt{2M}}{\lambda-\lambda_0},
\end{align}
where $\lambda_0$ is an integration constant. (See Appendix~\ref{app:geodesic}.)
In this case, $\Omega$ converges to zero in the limit $\lambda\to \lambda_0$ corresponding to a bifurcation $(n-2)$-sphere $v\to -\infty$, on which the Killing vector generating staticity vanishes~\cite{Townsend:1997ku}.
Our results are summarized in Table~\ref{table:results0}.
\begin{table*}[htb]
\begin{center}
\caption{The standard energy conditions that the mater field of the Semiz class-I solution with $k=1$ satisfies in different domains of $y$. $\Pi>0$ is required for $n\ge 5$ and $y>0$ is required for $n\ge 6$ in addition.}
\label{table:results0}
\begin{tabular}{|c|c|c|c|c|}\hline
Domain of $y$ & $n$ & Matter & $\zeta>0$ & $\zeta<0$ \\ \hline
$y\Pi>0$ & $n\ge 4$ & Perfect fluid & All & None \\ \hline
$y=0$ & $n=4$ & Vacuum & All & All \\ \cline{2-5}
& $n=5$ & Null dust & All & None \\ \hline
$y\Pi<0$ & $n=4,5$ & Anisotropic fluid & None & NEC \& SEC \\ \hline
\end{tabular}
\end{center}
\end{table*}

\subsubsection{Regular attachment to Schwarzschild-Tangherlini spacetime}

Lastly, we show that, for $n=4$ and $5$, two Semiz class-I spacetimes with the same $M$ but different $\zeta$ can be attached at the Killing horizon $y=0$ in a regular manner.
In other words, they can be attached without a massive lightlike thin-shell, which is a localized matter field at $y=0$ described by the delta-function.
As a special case, a Semiz class-I spacetime with $M>0$ can be attached to the Schwarzschild-Tangherlini vacuum black-hole spacetime at $y=0$.
To prove this, we will use the junction conditions at a null hypersurface developed in~\cite{Barrabes:1991ng,Poisson:2002nv},
(See also Section 3.11 in the textbook~\cite{Poissonbook}.)
Attachment of spacetimes at a Killing horizon has recently been studied in a more general setup in~\cite{Manzano:2022xct}.

For our purpose, we use the Semiz class-I metric (\ref{Semiz-I-v}) with the null coordinate $v$.
Let $\Sigma$ be a Killing horizon, which is a null hypersurface given by $y=0$.
The parametric equations $x^\mu=x^\mu({\eta},\theta^i)$ describing $\Sigma$ are $v=\eta$, $y=0$, and $\theta^i=z^i$.
The line element on $\Sigma$ is $(n-2)$-dimensional and given by
\begin{align}
\D s_{\Sigma}^2=h_{ab}\D w^a \D w^b=(2M)^{2/(n-3)}\gamma_{ij}\D z^i\D z^j(=\sigma_{ij}\D \theta^i \D 
\theta^j),\label{hab}
\end{align}
where $w^a=(\eta,z^i)$ is a set of coordinates on $\Sigma$.
Using them, we obtain the tangent vectors of $\Sigma$ defined by 
$e^\mu_a := \partial x^\mu/\partial y^a$ as
\begin{align}
e^\mu_{\eta}\frac{\partial}{\partial x^\mu}=\frac{\partial}{\partial v},\qquad e^\mu_i\frac{\partial}{\partial 
x^\mu}=\frac{\partial}{\partial z^i}.
\end{align}
An auxiliary null vector $N^\mu$ given by
\begin{align}
N^\mu \frac{\partial}{\partial x^\mu}=-(n-3)\Pi(y)^{(n-4)/(n-3)}\frac{\partial}{\partial y} \label{N-attachment}
\end{align}
completes the basis.
The expression $N_\mu \D x^\mu=-\D v$ shows $N_\mu N^\mu =0$, $N_\mu e^\mu_{\eta}=-1$, and $N_\mu e^\mu_i=0$.
The completeness relation of the basis on $\Sigma$ is given as
\begin{eqnarray}
g^{\mu\nu}|_\Sigma=-k^\mu N^\nu-N^\mu k^\nu+\sigma^{ij} e^\mu_{i}e^\nu_{j}, \label{comp}
\end{eqnarray}
where $k^\mu\equiv e^\mu_{\eta}$ and $\sigma^{ij}$ is the inverse of $\sigma_{ij}$.

Since $\sigma_{ij}$, $e^\mu_a$, and $N^\mu$ do not include $\zeta$ on $\Sigma$, $[e^\mu_a]=[N^\mu]=[\sigma_{ij}]=0$ are realized on $\Sigma$ when two Semiz class-I spacetimes with the same $M$ but different $\zeta$ are attached at $\Sigma$.
Here $[X]$ is the difference of $X$ evaluated on the two sides of $\Sigma$.
As a result, if the transverse derivative of the metric is continuous at $\Sigma$, a $C^1$ regular attachment of two spacetimes at $\Sigma$ is achieved.
In other words, there is no massive lightlike thin-shell at $\Sigma$ if $N^\rho[\partial_\rho g_{\mu\nu}]=0$ holds there.
This condition is equivalent to $[C_{ab}]=0$, where $C_{ab}:=(\nabla_\nu N_{\mu}) e^\mu_{a} e^\nu_b$ is the transverse curvature at $\Sigma$.

In the present case, nonvanishing component of $C_{ab}$ of $\Sigma$ are given by 
\begin{align}
\label{trans-C0}
\begin{aligned}
&C_{\eta\eta}=\Gamma^v_{vv}|_{y=0}=\frac{n-3}{2(2M)^{1/(n-3)}},\\
&C_{ij}=\Gamma^v_{ij}|_{y=0}=-(2M)^{1/(n-3)}\gamma _{ij}.
\end{aligned}
\end{align}
Since Eq.~(\ref{trans-C0}) is characterized only by $M$ and does not contain $\zeta$, $[C_{ab}]=0$ is realized if one attaches two Semiz class-I spacetimes with the same $M$ but different values of $\zeta$ at the Killing horizon $y=0$.

\begin{figure}[htbp]
\begin{center}
\includegraphics[width=0.9\linewidth]{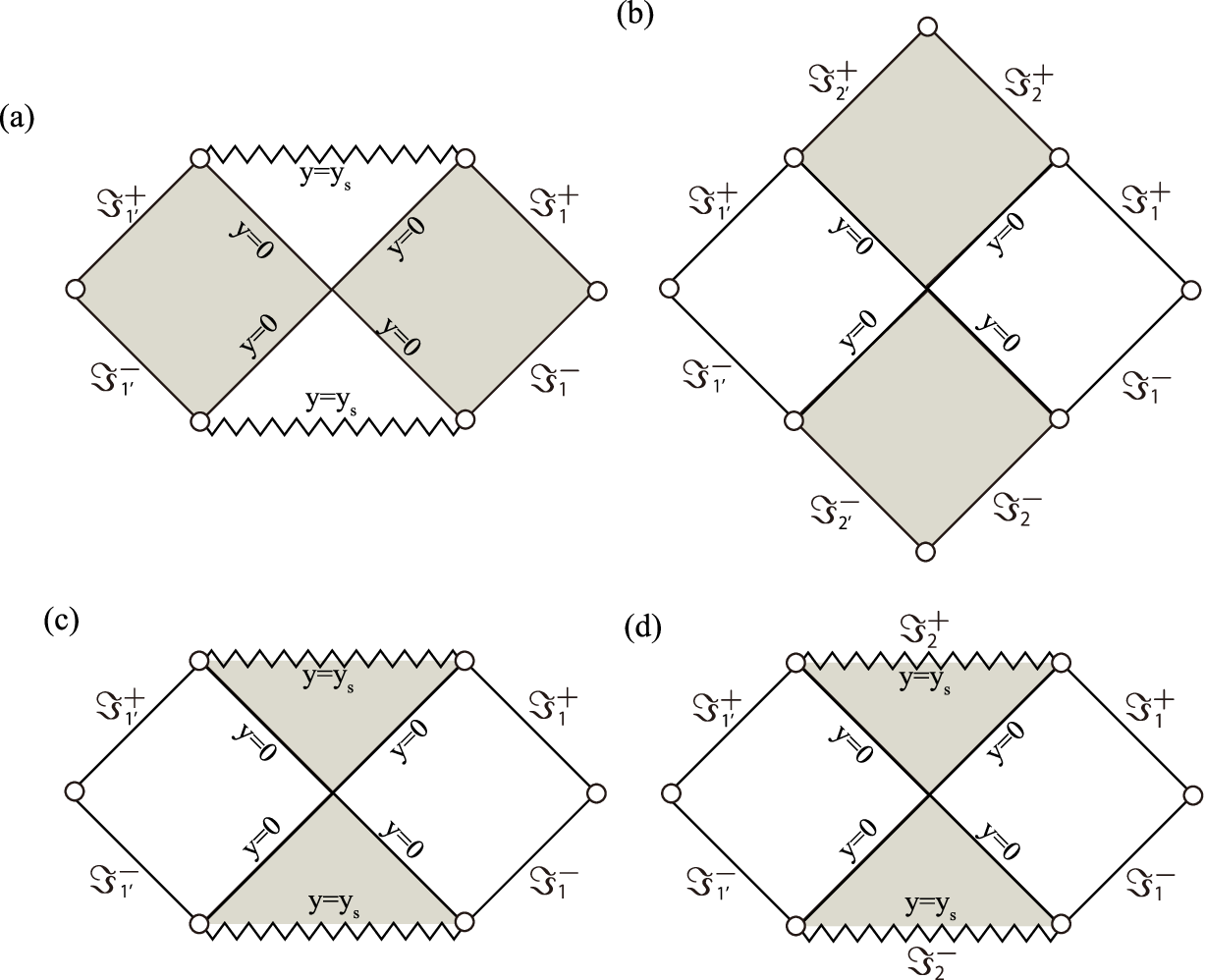}
\caption{\label{Fig:Semiz-BH} Possible Penrose diagrams of an asymptotically (locally) flat black hole in four and five dimensions constructed by attaching the Semiz class-I solution with $k=1$ to the Schwarzschild-Tangherlini solution at the Killing horizon $y=0$ (which is $r^{n-3}=2M$). Shaded regions are described by the Semiz class-I solution. See the caption of Fig.~\ref{Fig:Semiz-Penrose}.}
\end{center}
\end{figure}

As a special case, one can attach the Semiz class-I solution with $M>0$ in a regular manner to the Schwarzschild-Tangherlini black-hole spacetime described by Eq.~(\ref{Semiz-I-v}) with $\zeta=0$ and the same $M$.
In this construction, we assume that $y=0$ is a part of the Schwarzschild-Tangherlini spacetime so that a matter field is absent at the Killing horizon.
The Penrose diagrams of the resulting spacetime and the energy conditions that are fulfilled depending on the parameters are summarized in Table~\ref{table:results}.
Figure~\ref{Fig:Semiz-BH}(a) describes a perfect fluid hovering outside the Schwarzschild-Tangherlini black hole.
Figures~\ref{Fig:Semiz-BH}(b), (c), and (d) describe a Schwarzschild-Tangherlini black hole with its interior replaced by the Semiz class-I spacetime.
In particular, Fig.~\ref{Fig:Semiz-BH}(b) describes a non-singular black hole of the big-bounce type.
The difference between Figs.~\ref{Fig:Semiz-BH}(c) and (d) is the fact that the singularity of the latter is null infinity.
\begin{table*}[htb]
\begin{center}
\caption{Four- and five-dimensional asymptotically (locally) flat black holes constructed by attaching the spherically symmetric Semiz class-I solution with $M>0$ to the Schwarzschild-Tangherlini vacuum solution with the same $M$ at the Killing horizon $y=0$. The matter in the Semiz class-I region with $y>0$ and $y<0$ is a perfect fluid and an anisotropic fluid, respectively.}
\label{table:results}
\scalebox{0.9}{
\begin{tabular}{|c|c|c|c|c|}\hline
$n$ & $y<0$ & $y>0$ & Diagram in Fig.~\ref{Fig:Semiz-BH} & Energy conditions \\ \hline
$4$ & Schwarzschild & Semiz class-I ($\zeta<0$) & (a) for $M>0$ & None \\ \cline{2-5}
& Semiz class-I ($\zeta<0$) & Schwarzschild & (c) for $M>0$ & NEC \& SEC \\ \cline{2-5}
& Semiz class-I ($\zeta>0$) & Schwarzschild & (c) for $0<M\le M_{\rm ex(4+)}$ & None \\ 
& & & (b) for $M>M_{\rm ex(4+)}$ & \\ \cline{1-5}
$5$ & Tangherlini & Semiz class-I ($\zeta<0$) & (a) for $M>0$ & None \\ \cline{2-5}
& Semiz class-I ($\zeta<0$) & Tangherlini & (c) for $0<M< M_{\rm ex(5)}$ & NEC \& SEC \\ 
& & & (d) for $M= M_{\rm ex(5)}$ & \\ 
& & & (b) for $M>M_{\rm ex(5)}$ & \\ \cline{2-5}
& Semiz class-I ($\zeta>0$) & Tangherlini & (c) for $M>0$ & None \\ \hline
\end{tabular}
}
\end{center}
\end{table*}

We note that the configuration of Fig.~\ref{Fig:Semiz-BH}(a) does not contradict to the theorems by Shiromizu, Yamada, and Yoshino~\cite{Shiromizu:2006vh} for $n=4$ and by Rogatko~\cite{Rogatko:2012yq} for arbitrary $n(\ge 4)$, which prohibit any static configuration of a star composed of a perfect fluid in an asymptotically flat black-hole spacetime.
This is because the configuration of a static perfect fluid in Fig.~\ref{Fig:Semiz-BH}(a) is not a star.
Furthermore, it violates the dominant energy condition and the spacetime is not asymptotically flat but asymptotically locally flat, which are assumptions of the theorems in~\cite{Shiromizu:2006vh,Rogatko:2012yq}.

\section{Summary and concluding remarks}

We summarize the main results of the present paper.

\begin{itemize}
\item We have obtained the general $n(\ge 4)$-dimensional static solution with an $(n-2)$-dimensional Einstein base manifold for a perfect fluid obeying a linear equation of state $p=-(n-3)\rho/(n+1)$, which is a generalization of Semiz's four-dimensional general solution with spherical symmetry~\cite{Semiz:2020lxj} and consists of two classes of solutions.

\item The class-I and class-II solutions are dual to the topological Schwarzschild-Tangherlini-(A)dS solution and one of the $\Lambda$-vacuum direct-product solutions, respectively, through the $n(\ge 4)$-dimensional Buchdahl transformation in Proposition~\ref{prop:Buchdahl}.

\item The spherically symmetric class-I solution (\ref{Semiz-I-twopara-k=1}), characterized by two parameters $M$ and $\zeta$, is asymptotically locally flat as $r\to \pm\infty$.
While the metric and its inverse of the solution are $C^\infty$ at the Killing horizon $r^{n-3}=2M$ for $n=4$ and $5$, they are $C^1$ for $n\ge 6$ and then the Killing horizon turns to be a p.p. curvature singularity. The matter field on the Killing horizon is absent for $n=4$ and a null dust fluid for $n=5$. For $n=4$ and $5$, two spherically symmetric class-I spacetimes with the same $M$ but different $\zeta$ can be attached at the Killing horizon in a regular manner, namely without a lightlike massive thin-shell.

\item As a special case, a spherically symmetric class-I spacetime with $M>0$ and $\zeta\ne 0$ can be regularly attached at the Killing horizon to the Schwarzschild-Tangherlini vacuum black hole with the same $M$, which allows totally new configurations of an asymptotically (locally) flat black hole, as shown in Table~\ref{table:results}. 
\end{itemize}

In the last configuration of a black hole, as shown in Table~\ref{table:results}, all the standard energy conditions are violated when the static perfect fluid hovers outside a vacuum black hole.
On the other hand, the null and strong energy conditions can be satisfied when the dynamical region inside the event horizon of a vacuum black hole is replaced by the Semiz class-I solution with $\zeta<0$. 
In this case, the spacetime always involves a spacelike singularity inside the horizon for $n=4$ but it describes a non-singular black hole of the big-bounce type for $n=5$ if $M$ is larger than a critical value $M_{\rm ex(5)}:=-1/(8\zeta)(>0)$.
It should be emphasized that the exterior of this black hole is dynamically stable against linear perturbations because it is exactly the Schwarzschild-Tangherlini spacetime~\cite{Vishveshwara:1970cc,Ishibashi:2003ap}.
For the same reason, this black hole in four dimensions ($n=4$) cannot be distinguished from the Schwarzschild black hole by observations.
However, it is of course highly non-trivial how these new configurations of a black hole can be formed from, for example, gravitational collapse.

We have shown that, in the Semiz class-I solution for $k=1$, the metric on the Killing horizon is $C^\infty$ for $n=4$ and $5$, but it becomes $C^1$ for $n\ge 6$ and then the Killing horizon turns to be a p.p. curvature singularity. 
This is similar to the property of the $n(\ge 4)$-dimensional Majumdar-Papapetrou solution in the Einstein-Maxwell system~\cite{Majumdar:1947eu,Papaetrou:1947ib,Lemos:2005md}.
It describes a multi black hole with the $C^\infty$ metric at the extreme Killing horizon for $n=4$~\cite{Hartle:1972ya}, but the metric becomes $C^2$ at the horizon for $n=5$ and the Killing horizon turns to be a p.p. curvature singularity for $n\ge 6$~\cite{Candlish:2007fh}. (See also~\cite{Welch:1995dh}.)
Those examples show that the problem of differentiability of a Killing horizon in $n$-dimensional solutions with matter fields is an interesting problem worth pursuing.
Although a simple method to prove non-smoothness of a black-hole horizon has been proposed in~\cite{Kimura:2014uaa}, we still don't know the answer for static and spherically symmetric perfect-fluid solutions obeying a more general equation of state.
We leave this problem for further research.

\subsection*{Acknowledgements}
The author thanks Tetsuya Shiromizu for communications about the result in~\cite{Shiromizu:2006vh} and Tomohiro Harada for discussions on junction conditions.

\appendix

\section{Generalized Tolman-VI solution for $p=\chi\rho$}
\label{app:HW}

The Einstein equations (\ref{EFE-0}) with a perfect fluid (\ref{Tab-0}) obeying an equation of state $p=\chi\rho$ admits the following solution for $n\ge 4$ with $k\ne 0$:
\begin{align}
\label{MS-g}
\begin{aligned}
&\D s^2=-r^{4\chi/(1+\chi)}\D t^2+\frac{(n - 3)(1+\chi^2) + 2(n - 1)\chi}{k(n - 3)(1+\chi)^2}\D r^2+r^2\gamma_{ij}\D z^i\D z^j,\\
&u^\mu\frac{\partial}{\partial x^\mu}=r^{-2\chi/(1+\chi)}\frac{\partial}{\partial t},\qquad \rho=\frac{1}{\chi}p=\frac{2(n - 2)(n - 3)k\chi}{\kappa_n[(n - 3)(1+\chi^2) + 2(n - 1)\chi]r^2}.
\end{aligned} 
\end{align} 
The solution (\ref{MS-g}) with $n=4$ with $k=1$ is identical to the Tolman-VI solution with $B=0$~\cite{Tolman:1939jz} and a particular case of the self-similar (homothetic) static solution obtained by Henriksen and Wesson~\cite{HW1978}.

For the Lorentzian signature, $k[(n - 3)(1+\chi^2) + 2(n - 1)\chi]>0$ is required, which is equivalent to $\chi<\chi_-$ and $\chi>\chi_+$ for $k=1$ and $\chi_-<\chi<\chi_+$ for $k=-1$, where
\begin{align}
&\chi_\pm=\chi_\pm(n):=\frac{-(n-1)\pm 2\sqrt{n-2}}{n-3}(<0).
\end{align} 
$\chi_+(n)$ is a monotonically decreasing function and satisfies $\chi_+(4) =-3+2\sqrt{2}\simeq -0.17$ and $\lim_{n\to \infty}\chi_+(n)=-1$.
$\chi_-(n)$ is a monotonically increasing function and satisfies $\chi_-(4) =-3-2\sqrt{2}\simeq -5.8$ and $\lim_{n\to \infty}\chi_-(n)=-1$.
Therefore, $-1<\chi_+\le -3+2\sqrt{2}$ and $-3-2\sqrt{2}\le \chi_-<-1$ hold for $n\ge 4$.

It is noted that the solution~(\ref{MS-g}) is invariant under the Buchdahl transformation in Proposition~\ref{prop:Buchdahl} with $\Omega^{-2}=r^{4\chi/(1+\chi)}$.
The metric of the generated solution is 
\begin{align}
\D s^2=&-r^{-4\chi/(1+\chi)}\D t^2+\frac{(n - 3)(1+\chi^2) + 2(n - 1)\chi}{k(n - 3)(1+\chi)^2}r^{8\chi/[(n-3)(1+\chi)]}\D r^2 \nonumber \\
&~~~~~~~~~~~~~~+r^{2[(n-3)+(n+1)\chi]/[(n-3)(1+\chi)]}\gamma_{ij}\D z^i\D z^j,\label{HW-B}
\end{align} 
which becomes 
\begin{align}
&\D s^2=-{\bar r}^{4{\bar \chi}/(1+{\bar \chi})}\D t^2+\frac{(n - 3)(1+{\bar \chi}^2) + 2(n - 1){\bar \chi}}{k(n - 3)(1+{\bar \chi})^2}\D {\bar r}^2+{\bar r}^2\gamma_{ij}\D z^i\D z^j
\end{align} 
by the following coordinate transformation and a redefinition of the parameter:
\begin{align}
{\bar r}:=&r^{[(n-3)+(n+1)\chi]/[(n-3)(1+\chi)]},\\
{\bar\chi}:=&-\frac{(n-3)\chi}{(n-3)+2(n-1)\chi}.
\end{align}

\section{Radial null geodesics confined on the Killing horizon $y=0$}
\label{app:geodesic}

Consider a future-directed radial null geodesic $\gamma$ in the Semiz class-I spacetime with the metric (\ref{Semiz-I-v}), which is represented by $x^\mu=(v(\lambda),y(\lambda),0,\cdots,0)$ with its tangent vector $l^\mu=({\dot v},{\dot y},0,\cdots,0)$, where $\lambda$ is an affine parameter along $\gamma$ and a dot denotes differentiation with respect to $\lambda$.
Null geodesic equations $l^\nu\nabla_\nu l^\mu=0$ along $\gamma$ are written as
\begin{align}
0=&{\ddot v}-\frac{n-3}{2}\Pi^{-1/(n-3)}(y\Pi^{-1}\Pi'-1){\dot v}^2,\\
0=&{\ddot y}-\frac{(n-3)^2}{2}y\Pi^{-2/(n-3)}(y\Pi^{-1}\Pi'-1){\dot v}^2 \nonumber \\
&+(n-3)\Pi^{-1/(n-3)}(y\Pi^{-1}\Pi'-1){\dot v}{\dot y}-\frac{n-4}{n-3}\Pi^{-1}\Pi'{\dot y}^2,
\end{align} 
where a dot denotes differentiation with respect to $\lambda$.
Using $\Pi(0)=2M$ and $\Pi'(0)=1$, we show that the geodesic equations admit the following solution:
\begin{align}
v(\lambda)=&\frac{2(2M)^{1/(n-3)}}{n-3}\ln|\lambda-\lambda_0|+v_0,\qquad y(\lambda)=0,
\end{align} 
where $v_0$ and $\lambda_0$ are integration constants.
The above solution describes a radial null geodesic confined on the Killing horizon $y=0$ and $\lambda\to \lambda_0$ corresponds to a bifurcation $(n-2)$-sphere $v\to -\infty$, on which the Killing vector generating staticity vanishes~\cite{Townsend:1997ku}.


\end{document}